\documentclass[twocolumn,preprint]{aastex631}
\usepackage{threeparttablex}
\usepackage{longtable}
\usepackage{booktabs}
\usepackage{array}
\usepackage{makecell}

\def\h{~$h^{-1}$}
\def\m{~$\mu$m}
\def\R25{$R_{25}$}

\def\CIII {\ion{C}{3}]}

\def\HeII {\ion{He}{2}}

\begin{document}

\title{Merging Signatures in an Offset Lyman Continuum Emitter at Redshift 3.8}
\correspondingauthor{Fang-Ting Yuan}
\email{yuanft@shao.ac.cn}

\author{Fang-Ting Yuan}
\affiliation{Shanghai Astronomical Observatory, Chinese Academy of Sciences, 80 Nandan Road, Shanghai 200030, China}

\author{Zhen-Ya Zheng}
\affiliation{Shanghai Astronomical Observatory, Chinese Academy of Sciences, 80 Nandan Road, Shanghai 200030, China}

\author{Chunyan Jiang}
\affiliation{Shanghai Astronomical Observatory, Chinese Academy of Sciences, 80 Nandan Road, Shanghai 200030, China}

\author{Shuairu Zhu}
\affiliation{Shanghai Astronomical Observatory, Chinese Academy of Sciences, 80 Nandan Road, Shanghai 200030, China}
\affiliation{University of Chinese Academy of Sciences, No. 19A Yuquan
Road, Beijing 100049, China}

\author{Ruqiu Lin}
\affiliation{Shanghai Astronomical Observatory, Chinese Academy of Sciences, 80 Nandan Road, Shanghai 200030, China}
\affiliation{University of Chinese Academy of Sciences, No. 19A Yuquan
Road, Beijing 100049, China}

\author{Cheng Cheng}
\affiliation{Chinese Academy of Sciences South America Center for Astronomy, National Astronomical Observatories, CAS, Beijing 100101, China}

\begin{abstract}

Lyman continuum (LyC) emitters at $z>3$ provide critical samples for studying the contribution of galaxies to the ionizing background in the Epoch of Reionization. We collect a sample of $z>3$ LyC emitters, a dominant fraction ($\sim$60\%{--}70\%) of which shows spatial offsets between LyC emission and the non-ionizing continuum. From this sample, especially, we find a case of an offset LyC emitter, CDFS-6664 ($z=3.797$), which shows two components in the high-resolution Hubble Space Telescope (HST) and James Webb Space Telescope (JWST) images. The exceptionally rich data set of CDFS-6664 enables us to extract the two components across multiple wavelengths and estimate their physical properties. We show that CDFS-6664 is consistent with a major merger system with boosted star formation in both components and the offset LyC emission is most likely associated with the bluer and younger component in this merging system. Our result offers an example in which the offset can be caused by a merger. Future observations of more offset LyC emitters would elucidate the role that mergers play in the escape of LyC photons.

\end{abstract}


\section{Introduction} \label{sec:intro}

During the Epoch of Reionization (EoR), the neutral hydrogen dominating the Universe is ionized by radiation from the first luminous sources. As recent works have shown that radiation from quasars contributes insignificantly to cosmic reionization \citep{matsuoka2018,jiang2022,matsuoka2023}, galaxies become the most important sources that can provide the majority of ionizing photons for the reionization process. 

The contribution of galaxies to the ionizing budget depends on the escape fraction of ionizing photons which bears large uncertainties. 
It is not feasible to observe the escape of LyC photons from galaxies in the EoR due to the increase of intervening Lyman limit systems towards high redshifts. Many studies thus focus on detecting lower-redshift analogs. Star-forming galaxies at $3<z<4.5$ provide a very important sample as they are at the highest redshift where the LyC emission is observable. {In this redshift range, a few tens of galaxies with LyC emissions have been detected \citep{vanzella2015,shapley2016,micheva2017,fletcher2019,prichard2022,wang2023,kerutt2024,gupta2024}.  }

Among these galaxies, $\sim$ 60\%{--}70\% show positional offsets between the centers of their LyC and non-ionizing emissions\citep{iwata2009,mostardi2013,micheva2017,ji2020} (see Table \ref{tab:lyc_cands}). 
Except for a few possible cases of foreground contamination \citep{vanzella2010}, the physical mechanisms responsible for the offset LyC emission are still not well understood. Theoretically, offset LyC emission can be due to the non-isotropic distribution of neutral hydrogen in the HII regions and the non-uniform distribution of the HII regions in galaxies \citep{inoue2010,erb2015,verhamme2015,kakiichi2021}. Simulations have also found that the offset LyC emission can be caused by the unobscured young stars in the outskirts of galaxies \citep{gnedin2008}. However, it is difficult to explain large offsets (several kiloparsecs) simply by the clumpy distribution of ISM in host galaxies. These offset LyC emissions may provide additional constraints on LyC escaping mechanism, but it requires data with sufficient spatial resolution and wavelength coverage to understand their origins.

Here we report the case of CDFS-6664 (R.A.=03:32:33.326, Decl.=-27:50:7.363), which is a source at $z=3.797$ 
that has a LyC emission detection of about 0.2{\arcsec{}} (1.4 kpc) offset 
from the optical center \citep{yuan2021}. Among current detections of LyC emitters at $z>3$,{ CDFS-6664 has ultraviolet (UV) to near-infrared imaging and spectral data from Hubble Space Telescope (HST), James Webb Space Telescope (JWST), VLT/MUSE \citep{bacon2010}, and VLT/SINFONI \citep{eisenhauer2003} observations (Table \ref{tab:data}).} We identify this object as a merger system of two components from the high-resolution HST and JWST images. We examine the physical properties of this system by analyzing the spectral energy distributions (SEDs) of the two components extracted from UV to near-infrared images. Our results based on the SEDs indicate that the off-center LyC emission of CDFS-6664 is probably related to the galaxy merging process and most likely originates from one of the galaxies with a bluer UV spectral slope and younger age.

We present the observational data of CDFS-6664 in Section \ref{sec:data}. We describe our methods to analyze the properties of the components in Section \ref{sec:methods}. Our results are present in Section \ref{sec:results}. Finally, we discuss our findings in Section \ref{sec:discussion}.
Throughout this paper, we use proper distances. The AB magnitude system
$\mathrm{AB}=-2.5\log (f/\mathrm{Jy})+8.9$ and a cosmology of $\Omega_\mathrm{tot}$, $\Omega_\mathrm{M}$, $\Omega_{\Lambda}= 1.0, 0.3, 0.7$ with $H_{0} = 70~ \mathrm{km~s^{-1}~Mpc^{-1}}$ are used. 

\section{Data} \label{sec:data}

We identified CDFS-6664 as a LyC emitter candidate in a previous study according to a detection $\ga 5\sigma$ in the F336W band of HST image \citep[650{--}770 {\AA} rest-frame,][]{yuan2021}. The redshift of CDFS-6664 is secured to be 3.797 with the measurement of H$\beta$ and [OIII] emission lines in the VLT/SINFONI spectra \citep{maiolino2008}. The large [OIII]/[OII] ratio \citep[$>10$,][]{troncoso2014} and the double $Ly\alpha$ peaks shown in its spectrum further increase the credibility of this object being a LyC leaker. The multiwavelength photometric data and line ratios (e.g., \CIII~1909/\HeII~1640) show that this LyC source is more likely to be a stellar origin than AGN origin \citep{yuan2021}. In this work, we further investigate the properties of CDFS-6664 with the aids of the JWST data.

\subsection{Observations of CDFS-6664}
CDFS-6664 has rich observations, including high-resolution multiwavelength images and optical/near-infrared spectra (Table \ref{tab:data}). Further studies on the metallicity of this object have been made by \citet{sommariva2012} and \citet{troncoso2014}, showing that both the stellar metallicity and gas metallicity of CDFS-6664 is low ($12+\log(O/H)\approx 7.9$, about 0.2$Z_{\odot}$). The VLT/SINFONI integral field spectrum also provides dynamic information on this object. However, the study was not able to judge whether CDFS-6664 is a rotating disk or not \citep{gnerucci2011}. 

The redshift of this object has been further confirmed by the observation of the MUSE-Wide project \citep{urrutia2019}. The spectrum shows a significant Ly$\alpha$ emission line at the systemic redshift $z=3.797$. The Ly$\alpha$ line presents a double-peaked profile with a velocity difference $\Delta V\approx700$ km/s. The central escape fraction of Ly$\alpha$ emission within $\pm 100$ km/s of the systemic velocity $f_\mathrm{cen}$ is about $28$\%. 

For the LyC detection of CDFS-6664, we examine the following hypotheses of contamination. First, we argue that it is unlikely that the offset F336W band detection is due to contamination from an intervening star. The signal is only detected in the F336W band but any bluer (F275W) or redder bands (3D-HST and JWST bands) at exactly the same position, indicating a peak of SED distribution at around 3300\AA. The only possible star is a white dwarf with an effective temperature of about 15000K. However, according to the spectra of such a white dwarf, the magnitude in the F275W band should be brighter than 27.6 mag, meaning at the F275W band it should have a detection greater than $5\sigma$, contrary to the observation. Furthermore, such a star would imply a distance larger than 10 kpc. Since CDFS field is oriented at a considerable angular distance from the galactic disk, the chance of observing such a star on the line of sight of CDFS-6664 is very little. Therefore, it is unrealistic to consider the signal coming from any intervening star.

The LyC signal is also not likely to come from a foreground galaxy interloper. No contaminants are observed around this object at the LyC leaking position in the deep HST images and JWST images. The LyC signal is not likely to be due to a red-leak of the UVIS filter \citep{windhorst2011}, because the signal is offset from the optical center where the flux of the red band peaks.
Considering the spectral and imaging features, the signal detected at the F336W band of CDFS-6664 is most likely to be the LyC emission. 

\begin{deluxetable*}{lc}
\scriptsize
\tablecaption{Observations of CDFS-6664.}
   \label{tab:data}
\tablewidth{1.0\columnwidth}
\tablehead{
\colhead{Observations} & \colhead{Refs.}
}
\startdata
        Optical spectrum (Ly$\alpha$) & MUSE-Wide \citep{urrutia2019}\\
        IR spectrum ([OIII], [OII], H$\beta$) & AMAZE VLT/SINFONI \citep{maiolino2008}\\
        UV images (F275W, F336W) & the Hubble Deep UV Legacy Survey \citep[HDUV, v1.0][]{oesch2018} \\
        HST Optical images (F435W{--}F850LP) & \makecell{Hubble Legacy Fields Data (HLF, v2.0) \\ \citep[][]{https://doi.org/10.17909/t91019,illingworth2016,whitaker2019}} \\
        HST Optical/IR images (F435W{--}F160W) & \makecell{3D-HST (v4.0, science images) \\ \citep[][]{grogin2011,koekemoer2011,skelton2014,https://doi.org/10.17909/t9jw9z}}  \\
        JWST NIRCam IR images (F182M{--}F480M) & GO 1963, \citep{williams2023}, reduced by G. Brammer (2022) 
\enddata
\end{deluxetable*}

\subsection{Two components in CDFS-6664} \label{subsec:comps}
\begin{figure*}[ht]
    \centering
    \includegraphics[width=\linewidth]{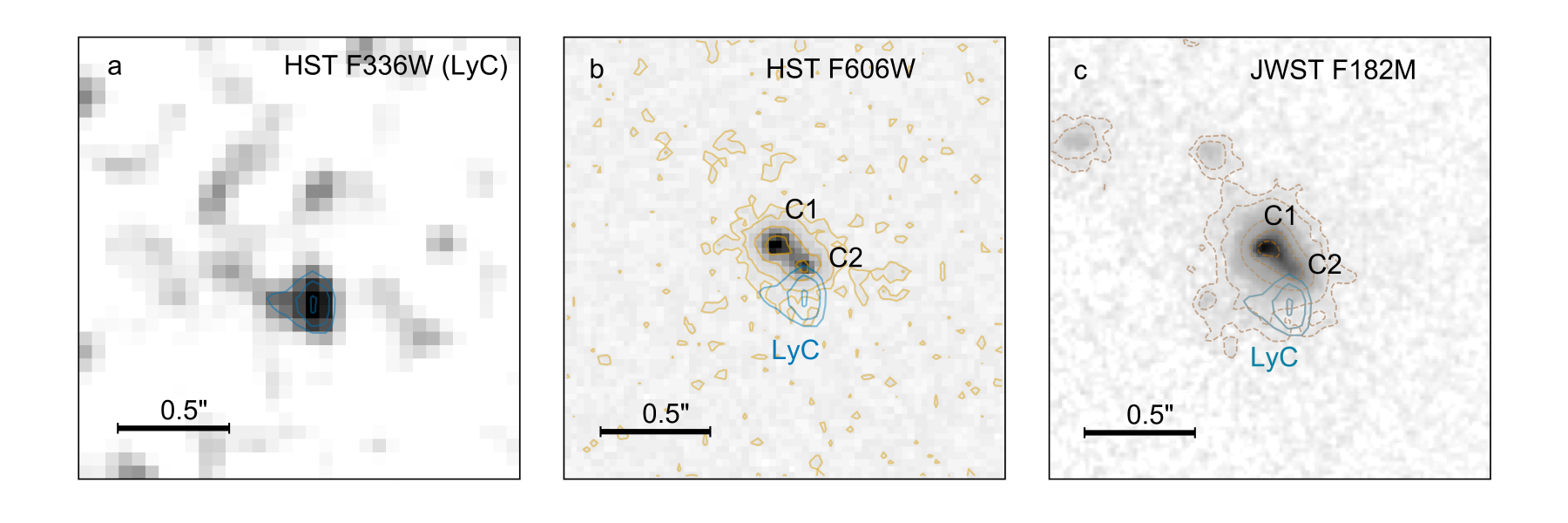}
    \caption{Multiwavelength images of the LyC emitter, CDFS-6664 ($z=3.797$). {Contours of the fluxes in LyC (blue, 3, 4, and 5$\sigma$), HST F606W (orange, 2, 5, 15, and 30$\sigma$), and JWST F182M (red dashed, 2, 5, 30, and 60$\sigma$) are also displayed.} \textbf{a}. Smoothed image of LyC emission in HST F336W band (rest-frame 650-770\AA{}). \textbf{b}. Two components (C1 and C2) of CDFS-6664 in HST F606W band image (rest-frame 1300\AA{}). C1 and C2 show comparable fluxes and sizes. \textbf{c}. Two components (C1 and C2) in JWST F182M band image (rest-frame 3800\AA{}). C2 becomes significantly fainter than C1, indicating C2 is the bluer component. The LyC emission is at the side of C2, implying a connection between C2 and LyC. According to the positions of stars in the field of CDFS-6664, these multiwavelength frames are aligned with a precision of less than 0.1\arcsec{}.}
    \label{fig:img}
\end{figure*}

Figure \ref{fig:img} shows the images of CDFS-6664 from HST and JWST observations \citep{skelton2014,oesch2018,williams2023}. The HST F336W image shows the LyC emission of CDFS-6664, while the HST F606W and JWST F182M images show the structure of the nonionizing emissions. It is evident from the F606W and F182M images that CDFS-6664 contains two main components, C1 and C2. The depth of the F182M image \citep[29.3 mag, 5$\sigma$][]{williams2023} reveals that the morphology is disturbed. The LyC emission is distributed at the side of the C2 component. By visually inspecting these images we find that C2 appears fainter but C1 remains relatively unchanged in the F182M band compared with their appearances in the F606W band, suggesting that C2 has a bluer color than C1. We present a quantitative analysis of these two components in Section \ref{sec:methods}.

\section{Methods} \label{sec:methods}

With the multiwavelength images of HST and JWST, we attempt to extract the components in CDFS-6664 and analyze their SEDs. In the optical bands (F435W to F850LP), we have examined images from both Hubble Legacy Fields Data \citep[HLF,][]{illingworth2016,whitaker2019} with a pixel scale 0.03\arcsec{} and the 3D-HST images with a pixel scale 0.06\arcsec{}. We present the results based on the 3D-HST images considering the consistency with redder bands (F125W and F160W). The photometry difference between HLF and 3D-HST is less than 0.1 mag and does not affect our results. {For HST bands, we use the PSF in each band provided by 3D-HST. For JWST bands, we adopt theoretical PSF models generated from WebbPSF \citep{perrin2012,perrin2014}.}

\subsection{Decomposing images using GALFIT}

\begin{figure}
    \centering
    \includegraphics[width=\linewidth]{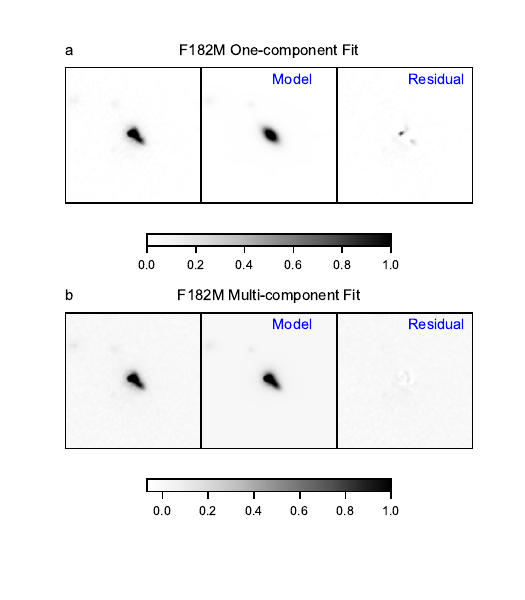}
    \caption{GALFIT results for JWST F182M image of CDFS-6664 in $2\arcsec{}\times2\arcsec{}$ boxes. {All the images are normalized to the maximum of the observed F182M image.} \textbf{a}. Results of fitting with one S\'{e}rsic component. The image can not be fitted well, with prominent structures left in the residual image. \textbf{b}. Results of fitting with two S\'{e}rsic and a PSF. The quality of the fitting is improved with this model. }
    \label{fig:f182galfit}
\end{figure}

{
The decomposition process utilizes the JWST F182M image, selected for its highest spatial resolution and the deepest imaging among all available bands. We find that using a single component for CDFS-6664 in the F182M band results in a poor fit to the image, as indicated by noticeable residuals (as shown in Figure \ref{fig:f182galfit}a). We then increase the number of components until no apparent residuals remain (Figure \ref{fig:f182galfit}b). Furthermore, we incorporate constraints on the effective radii of these components, derived from the HST F606W image. 

Our refined fitting approach indicates that C1 is best modeled with a combination of a S\'{e}rsic profile and a PSF, while C2 is adequately described by a S\'{e}rsic profile. A summary of the morphology models of the two S\'{e}rsic components is presented in Table \ref{tab:components}. The presence of a PSF component in C1 may indicate a certain level of nuclear activity. 

We also estimate the errors for the morphological parameters. We adopt an approach similar to \citet{lange2016}. We examine the variations of the output parameters using 162 sets of initial parameters of $R_e$ and $n$. We set the initial $R_e$ to range from 2 to 6 pixels for C1, and from 1 to 5 pixels for C2. The initial $n$ ranges from 0.5 to 1.5 for C1 and 0.3 to 1 for C2. The results exhibit only slight variations. The $1\sigma$ relative error for each morphological parameter is less than 0.001.}

\begin{figure*}
    \centering
    \includegraphics[width=0.8\linewidth]{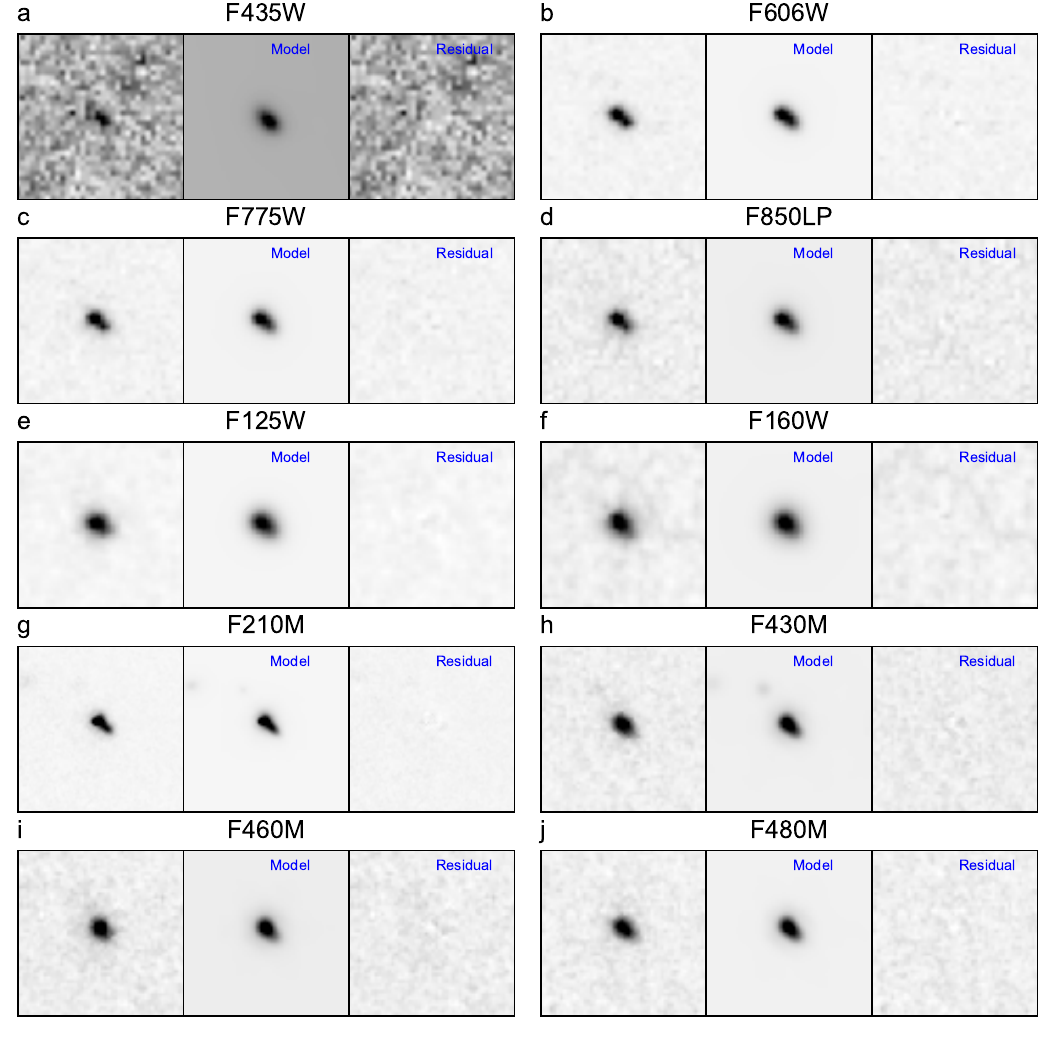}
    \caption{GALFIT decomposition for image of CDFS-6664 in each band based on the model derived from JWST F182M image.  \textbf{a} to \textbf{j}: The left is the observed image, the middle is the best model image by GALFIT, and the right is the residual image. The residual images show that the models can fit images in each band well.}
    \label{fig:galfit}
\end{figure*}

{
The models derived from the F182M image can represent the intrinsic light profile. Assuming a consistent intrinsic light profile across all bands, we then apply these models to fit images of other bands and extract the flux associated with each model. In each band, the models are convolved with respective PSF to take into account the inhomogeneous spatial resolution. 

During the fitting process, we keep the parameters of the intrinsic light profile, including the effective radius ($R_e$), S\'{e}rsic index ($n$), axis ratio ($b/a$), and position angle (P.A.), as fixed parameters. We also maintain the relative positions of the components consistent across bands but permit slight adjustments to their absolute positions. This flexibility accommodates potential astrometric discrepancies between HST and JWST data, which our star position comparisons in this region indicate to be less than 0.1\arcsec{}. The models can fit images in each band well, as shown in Figure \ref{fig:galfit}. The resulting magnitudes of the models are used for SED fitting.
}

\begin{table}
    \centering
    \begin{tabular}{l l l}
    \hline
        Parameters &  C1          &  C2 \\
    \hline
        Effective radius $R_e$      &  0.08\arcsec{} &  0.08\arcsec{} \\
        S\'{e}rsic index $n$ &  1.3         &  0.5        \\
        Axis ratio $b/a$      &  0.9         &  0.5        \\
        Position angle       &  $6^{\circ}$ &  $45^{\circ}$    \\
    \hline
    \end{tabular}
    \caption{Morphological parameters of the two S\'{e}rsic components extracted by GALFIT.}
    \label{tab:components}
\end{table}

\subsection{SED fitting}
\label{subsec:sedfitting}

We obtain the fluxes of C1 and C2 in each photometric band based on the integrated magnitude of the models derived from GALFIT. The SED of C1 is computed by summing the fluxes of its S\'{e}rsic and PSF components. The PSF component is about two times weaker in total than the S\'{e}rsic component of C1 and the SED is bluer. Therefore, including or excluding the PSF component in the SED of C1 does not affect our conclusion on the difference between the properties of C1 and C2. 

We use Code Investigating GALaxy Emission \citep[CIGALE][]{burgarella2005,noll2009,boquien2019} to fit the SED with models of stars, gas, and dust.  
We assume the Salpeter initial mass function \citep[IMF,][]{salpeter1955} and adopt the stellar population models from \citet{bc2003} (BC03). The star formation history (SFH) is assumed to follow the delayed form: SFR$\propto t/\tau^2 e^{-t/\tau}$, where $t$ is the age from the onset of star formation, and $\tau$ is the star formation time-scale. 

The nebular emission is modeled based on the templates of \citet{inoue2011b} using the CLOUDY code. The input parameters include the metallicity, radiation strength $U$, the dust attenuation, and the escape fraction of the Lyman continuum photons, $f_\mathrm{esc}$. The metallicity is assumed to be the same as that of stars. The radiation strength $U$ is fixed to $\log U=-2.5$. 

For the effect of the dust, we assume that the dust obscuration of the stellar part obeys the Calzetti Law \citep{calzetti2000} modified by adjusting the slope \citep{noll2009}:
\begin{equation}
    A(\lambda)=E(B-V)_\mathrm{star}k'(\lambda)\left(\frac{\lambda}{\lambda_\mathrm{V}}\right)^\delta,
\end{equation}
where $k'(\lambda)$ is from \citet{calzetti2000}, and $\delta$ indicates the modification to the slope. For the nebular emission, the dust obscuration is assumed to follow a simple screen model and a Milky Way extinction curve \citep{cardelli1989}. We assume that the dust does not affect the LyC emission. This assumption is valid when the LyC emission is escaped from transparent paths. 

The SED fitting includes fluxes measured from HST and JWST images (HST F435W, F606W, F775W, F850LP, F125W, F160W bands, JWST F182M, F210M, F430M, F460M, F480M bands). Since the LyC band (F336W) may deviate significantly from the average IGM model used in CIGALE \citep{meiksin2006}, we do not include it in the SED fitting, but we confirm that the inclusion of these bands does not affect the fitting results significantly. 

The output parameters are estimated through a Bayesian approach. For each model, the $\chi^2$ is calculated by comparing the model fluxes and the observed fluxes. The probability distribution function (PDF) of each parameter is then calculated by using the summation of the probability of each bin of parameter space to construct the PDF. Then the parameters and the corresponding uncertainties are estimated from the PDF by taking the likelihood-weighted mean and the standard deviation of all models.

The global quality of the fit is estimated using a reduced $\chi^2$ of the best model (that is, the model with the smallest $\chi^2$), $\chi^2_r=\chi^2/(N-1)$, where $N$ is the number of input fluxes. Both the SEDs of C1 and C2 can be fitted well, with $\chi^2_r<2.0$.

The models and input parameters used in the fitting are summarized in Table \ref{tab:sedpar}. In the fitting, we assume the parameters of dust attenuation ($E(B-V)_\mathrm{lines}$ and $\delta$) to be the value obtained from the SED fitting of the CDFS-6664 as one object. We note that the dust attenuation and the age of the stellar population are degenerated. In this paper, we assume that C1 and C2 have similar dust attenuation, and therefore attribute the difference between the SEDs of C1 and C2 to the age of the stellar population. 
{When we allow the attenuation parameters to vary freely, we observe that the overall results remain largely unchanged. However, this freedom introduces greater uncertainties in the output parameters. Our findings indicate that the ages of C1 and C2 approach each other, with C1 at $13\pm9$ Myr and C2 at $6\pm4$ Myr. Despite this, C2 remains the younger of the two. The Star Formation Rate (SFR) for C1 is $24.0\pm11.1$ $M_\mathrm{\odot}$yr$^{-1}$, while for C2 it is $46.1\pm34.9$ $M_\mathrm{\odot}$yr$^{-1}$.
The UV-continuum slope ($\beta$) for C1 is $-1.9\pm0.1$, and for C2 is $-2.5\pm0.1$.
Lastly, the Far Ultraviolet (FUV) attenuation ($A_\mathrm{FUV}$) of C1 ($1.0\pm0.1$ mag) is found to be greater than that of C2 ($0.8\pm0.1$ mag).
}  

\subsection{Escape Fraction of Ionizing Photons}
\label{subsec:fesc}

{
In our SED fitting process for C1 and C2, the escape fraction of ionizing photons, $f_\mathrm{esc}$, is treated as a free parameter within the nebular emission model, allowing it to be directly calculated from the fit results. 
However, $f_\mathrm{esc}$ derived in this way has large uncertainties due to the lack of constraints from Balmer line data and LyC data in the fitting. 

We verify the results using another approach following \citet{steidel2001}:
\begin{equation}
    f_\mathrm{esc} = \frac{(F_\mathrm{UV}/F_\mathrm{LyC})_\mathrm{int}}{(F_\mathrm{UV}/F_\mathrm{LyC})_\mathrm{obs}}\cdot \frac{1}{T_\mathrm{IGM}}\cdot \exp{^{-\tau_\mathrm{UV, dust}}},
\end{equation}
where $T_\mathrm{IGM}$ is the transmission of the IGM. The observed LyC flux can be derived from the F336W band image. We use the total magnitude (see the details of the measurement in \citealt{yuan2021}). The observed UV flux is estimated by interpolating the decomposed SED to 1500 \AA. The intrinsic UV-to-LyC ratio, $(F_\mathrm{UV}/F_\mathrm{LyC})_\mathrm{int}$ is determined from the best model of the SED fitting. Dust attenuation, $\tau_\mathrm{UV, dust}$, is calculated using the derived color excess $E(B-V)$ and the Calzetti attenuation curve. According to \citet{steidel2018}, the $T_\mathrm{IGM}$ is $<0.35$ for 99\% line-of-sight at $z=3.5$. Assuming the observed LyC emission is all from C2, we can then obtain a lower limit of $f_\mathrm{esc}$ for C2 and compare the result with the output of SED fitting.
}

\begin{table*}
\centering
\caption{Parameters in SED fitting. \label{tab:sedpar}}
\begin{tabular}{lc}
\hline
Star formation history, delayed form& \\
\hline
Time scale ($\tau$) & 10\\
Age & 2, 5, 10, 20, 50, 100, 200, 500, 1000, 1500\\
\hline
Stellar population \citep{bc2003} & \\
\hline
IMF & \citep{salpeter1955} \\
Metallicity & 0.0001, 0.0004, 0.004 ($Z_{\odot}=0.02$) \\
\hline
Nebular \citep{inoue2011b} & \\
\hline
LyC escape fraction ($f_\mathrm{esc}$) & 0.0, 0.1, 0.2, 0.3, 0.4, 0.5, 0.6, 0.7, 0.8, 0.9, 1.0 \\
Radiation strength ($\log U$) & -2.5 \\
\hline
Dust attenuation (modified Calzetti dust attenuation curve) & \\
\hline
$E(B-V)_\mathrm{lines}$ & 0.075\\
Modifying the attenuation curve ($\delta$) & -0.4\\
$E(B-V)$ factor ($f_\mathrm{ebv}$) & 1.0\\
\hline
\end{tabular}
\end{table*}

\section{Results}
\label{sec:results}

\begin{figure*}[ht!]
    \centering
    \includegraphics[width=0.95\linewidth]{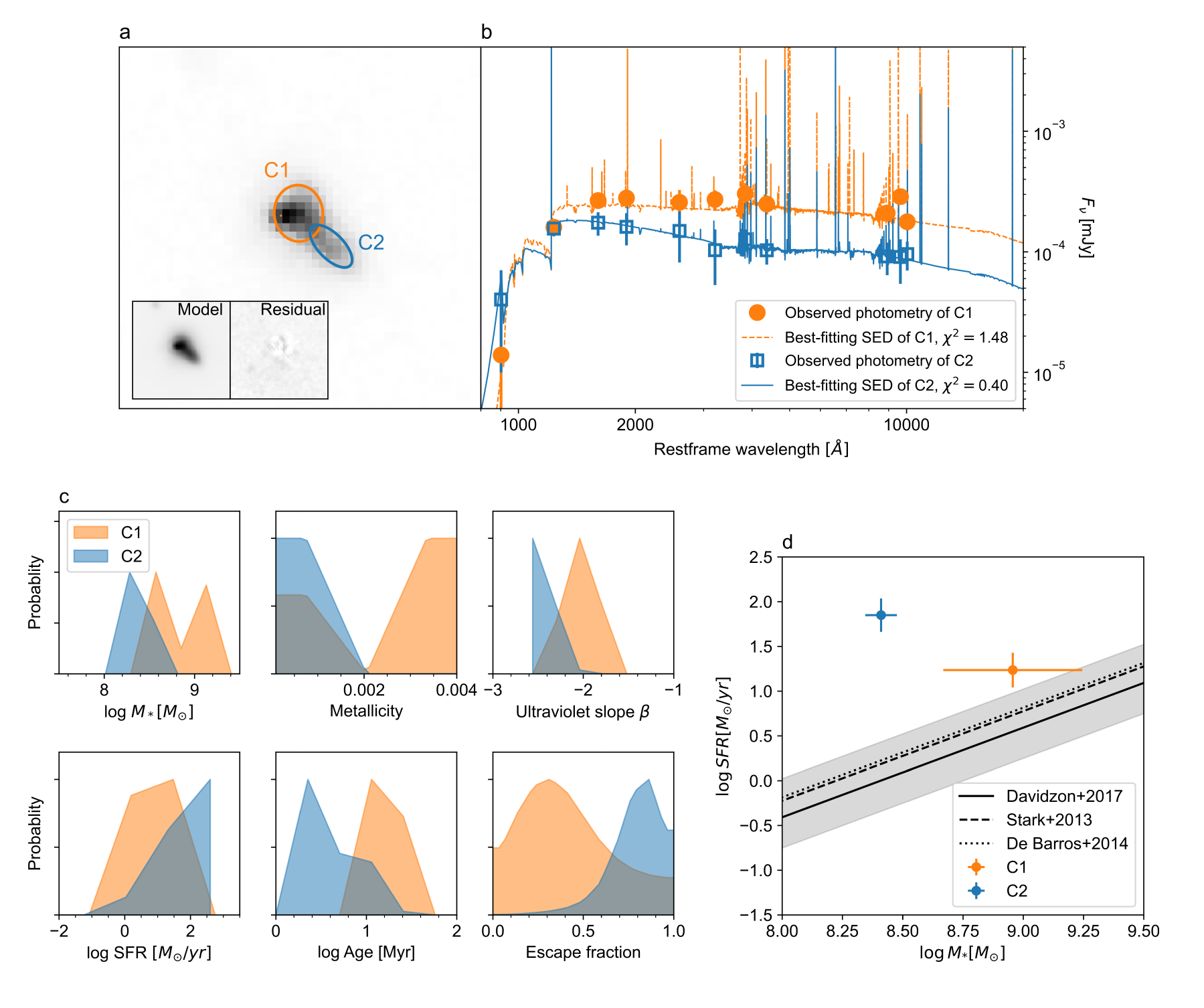}
    \caption{Physical properties of CDFS-6664's two components. \textbf{a}. Effective radii of C1 (orange) and C2 (blue) extracted by GALFIT based on the JWST F182M image. The model and residual images are shown on the lower left of the image. \textbf{b}. SEDs and best-fitting models of the two components C1~(orange) and C2 (blue). The models can fit the SEDs well, with a goodness of fit $\chi^2<2$. \textbf{c}.~Probability~density distributions for physical parameters derived from SED fitting for C1 (orange) and C2 (blue). C1 and C2 are consistent with a major merger system with a stellar mass ratio $M_\mathrm{*, C1}/M_\mathrm{*,C2}$ of $3.5$. C2 has a bluer ultraviolet slope and younger stellar populations than C1. \textbf{d}. Star formation rates of C1 and C2 compared to the main sequence of star-forming galaxies at redshift $\sim$4. The shaded area indicates the $3\sigma$ uncertainty. C1 and C2 lie significantly above the main sequence, indicating that starbursts may be triggered by merging. }
    \label{fig:pdf}
\end{figure*}

Taking advantage of the rich multi-wavelength datasets, we are able to extract the morphological information of the two galaxies using the two-dimensional fitting algorithm. Our modeling is based on the F182M image which is the sharpest and deepest. The model and residual images are shown in Figure~\ref{fig:pdf}\textbf{a}. 
The projected separation of C1 and C2 is 0.13\arcsec{} (0.92 kpc). The peak position of the LyC emission is located at 0.23\arcsec{} and 0.12\arcsec{} to the centers of C1 and C2, respectively.  

C1 and C2 are unlikely to be clumpy structures. The S\'{e}rsic index of C1 is 1.3, indicating a disk structure. The S\'{e}rsic index of C2 is 0.5 and the axis ratio ($b/a$) is 0.5, indicating that C2 is an inclined non-spherical component. The effective radius of each component is about 0.6 kpc, consistent with the size of galaxies with similar UV luminosities observed at redshift 4 \citep[$\sim$ 0.75 kpc,][]{shibuya2015}. In each band, C1 and C2 contribute to more than 75\% flux of CDFS-6664. C1 and C2 contribute to about 85\% stellar mass of CDFS-6664. As shown in \citet{guo2012}, clumps can only contribute 20\% optical luminosity and stellar mass of a galaxy. Therefore, C1 and C2 are more likely to be individual galaxies than star-forming clumps. 

{Several pieces of evidence suggest that C1 and C2 are engaged in an interaction. The probability that C1 and C2 are superposed by mere chance is quite low.} If we adopt the surface number density of objects according to \citet{skelton2014} ($\sim3.0\times10^5$mag$^{-1}$deg$^{-2}$), the probability that C2 is overlapped by C1 about 0.1\%. Furthermore, We extract the SEDs of C1 and C2 based on the morphological models (Figure~\ref{fig:pdf}\textbf{b}). The photometric redshifts estimated from the SEDs of C1 and C2 are 3.9$\pm$0.2 and 3.8$\pm$0.1, consistent with the spectroscopic redshift, {which further supports the hypothesis that C1 and C2 are in the process of merging. The extended features observed around the components in the JWST images provide additional indications of an ongoing interaction between C1 and C2.}

We also find evidence of merging from the Ly$\alpha$ emission line based on the spectrum observed by the MUSE Wide project \citep{urrutia2019} with a resolution $R\sim 3000$. The Ly$\alpha$ profile shows a strong, asymmetric main peak at restframe 1216\AA{} and a weaker peak at 1214\AA{} \citep{yuan2021}. The weak peak can be naturally explained by scattering in the expanding shell of the neutral hydrogen surrounding the Ly$\alpha$ emitting region \citep{yamada2012}.
{In addition to the prominent double-peaked structure, we observe that the strong peak (at restframe 1216\AA{}) itself can be well-fitted by two Gaussian curves, potentially suggesting the presence of a merging system. However, we note that the Ly$\alpha$ profile of a single galaxy can be much more complicated than a Gaussian profile, due to the processes such as multiple scattering and outflows.} Further investigation requires integral field spectra with higher spatial and spectral resolution.

\begin{deluxetable}{ccccc}
\tablecaption{Physical properties of C1 and C2 derived from SED fitting.}\label{tab:props}

\tablehead{\colhead{} & \colhead{C1} & \colhead{C2} } 

\startdata
$\log M_{*}/M_{\odot}$ & 8.96$\pm$0.29 & 8.41$\pm$0.07 \\
$\mathrm{SFR}$ [$M_{\odot}$yr$^{-1}$] & 17.7$\pm$7.7 & 70.7$\pm$30.4 \\
Age [Myr] & 4$\pm$3 & 17$\pm$11 \\
$f_\mathrm{esc}$ & 0.4$\pm$0.3 & 0.8$\pm$0.2 \\
$\beta$ & -2.1$\pm$0.1 & -2.4$\pm$0.1 \\
$M_\mathrm{UV}$ [mag] & -20.56$\pm$0.05 & -20.17$\pm$0.07 \\
\enddata

\end{deluxetable}

{We further investigate the physical properties of C1 and C2, respectively. Figure~\ref{fig:pdf}\textbf{c} shows the probability distribution of the physical parameters of C1 and C2. The values derived from the SED fitting are listed in Table \ref{tab:props}.} The results show that C1 and C2 would form a major merger system with a stellar mass ratio $M_\mathrm{*, C1}/M_\mathrm{*,C2}$ of $3.5$. The UV absolute magnitude ($M_\mathrm{UV}$) is -20.56 mag for C1 and -20.17 mag for C2. The star formation in C1 and C2 is vigorous, with the logarithmic specific star formation rates ($\log$ sSFRs/yr$^{-1}$) of $-7.7\pm0.8$ and $-6.6\pm0.5$, respectively. These \mbox{sSFRs} are significantly higher than the average sSFR ($\log$ sSFRs/yr$^{-1}$$\sim$$-8.4$) of galaxies at redshift around 4 \citep{stark2013,debarros2014,davidzon2017,popesso2023} (Figure~\ref{fig:pdf}\textbf{d}), suggesting starbursts triggered by merging. The ongoing extreme starbursts may cause intense LyC emissions.

The properties of the stellar populations in C1 and C2 are different. C2 has stellar populations with a mass-weighted age of about $4\pm3$ Myr which is younger than the stellar populations in C1 with an age of about $17\pm11$ Myr. The ages are consistent with the scheme for LyC escape proposed by \citet{naidu2022} that C2 is in a phase ($\sim2-10$ Myr) characterized by strong feedback from the early formed stars and a high LyC escape fraction while C1 is in a phase ($\sim10-100$ Myr) that feedback weakens, dust created and gas thickens where LyC cannot easily escape from the galaxy. C2 also has a bluer UV slope ($\beta$) compared to C1, which is consistent with a high LyC escape fraction \citep{chisholm2022,lopez2023}. 

Furthermore, C2 is more actively star-forming than C1. We estimate the ionizing photon production rate $Q_\mathrm{H0}$ to be $1.4\times10^{54}$s$^{-1}$ for C1 and $7.9\times10^{54}$s$^{-1}$ for C2 based on the best fitting SED models. The ionizing photon production efficiency is 24.83 erg$^{-1}$Hz and 25.82 erg$^{-1}$Hz for C1 and C2, respectively. Therefore, C2 can produce ionizing photons more effectively than C1. 

{
All the physical properties suggest that C2 is more likely to be responsible for the LyC emission. The derived $f_\mathrm{esc}$ from SED fitting is $41\pm27\%$ for C1 and $81\pm16\%$ for C2. Assuming that C2 contributes all the observed LyC emission in the F336W band, we obtain that $f_\mathrm{esc}$ of C2 is $>37$\%, consistent with the results given by the SED fitting. Such a high $f_\mathrm{esc}$ is significantly larger than the value required to complete the cosmic reionization \citep[e.g.,][]{madau1999,ouchi2009,wise2009,finkelstein2012,finkelstein2019,robertson2015,bera2023}.
}

\section{Discussion}
\label{sec:discussion}
We find that about 60\% of the LyC emitter candidates detected at $z>3$ have positional offsets between the LyC and the nonionizing emissions in the reference bands. Nevertheless, there has been no clear evidence supporting previous scenarios explaining the offset, including shocks, substructures, or infalling of a minor component \citep{iwata2009,inoue2011,micheva2017}. The case of CDFS-6664 provides evidence that the offset LyC emission may be connected to galaxy merging. 

\subsection{Candidate LyC emitters at z$>$3: More than half show spatial offsets of LyC}
\label{sec:lyc_collection}
We collected the LyC detections at $z>3$ in previous works and examined whether their LyC emissions are offset from the center of the reference band based on the available images. Cross-matching these works, we remove known AGNs, duplicated detections and possible contaminants from each work. We keep only the most credible candidates that have spectroscopically confirmed redshifts. The resulting sample includes {100} candidate LyC emitters (Table \ref{tab:lyc_cands}). 

In this sample, {89} galaxies have images in both the LyC and reference non-ionizing bands. {For these galaxies, we determine the existence of an offset in the LyC emission by comparing the peak positions in the LyC band and the reference band images. Several studies have already quantified the distance between the LyC band and the reference band \citep[e.g.,][]{nestor2013,ji2020,gupta2024}, and we directly adopt their measurements. We consider an offset to be valid if the distance is at least 0.1\arcsec. For ground-based observations, we require greater distance considering that the astrometric error can exceed 0.1\arcsec. That is, the distance must exceed the astrometric error, with a minimum requirement of 0.1\arcsec. For example, the astrometric error of Subaru/Suprime-Cam in the NB359 band is about 0.2\arcsec \citep{micheva2017}. In NB359 case, we require a distance more than 0.2\arcsec{} to be an offset.} 
In our study, an offset is only identified when there is a high degree of certainty. Any measurements that do not meet this stringent criterion are not counted as offsets. In four galaxies, the LyC signal is too faint to ascertain the presence of an offset. We categorize the offset as undetermined, denoted by a ‘-’ in Table \ref{tab:lyc_cands}.

We identified offsets between the LyC and non-ionizing emissions from 61 of these {89} galaxies. One may argue that some offset LyC detections are due to foreground contaminants. However, even if we count half of these offset LyC detections as foreground contaminants \citep{nestor2013}, there would be more than half (30/{58}) of LyC emitters with offset LyC emission. Furthermore, {29} LyC emitters in this sample have been observed in both the LyC and reference bands by HST, which provides images with significantly higher spatial resolution than ground-based facilities. About 18 of these {29} LyC emitters ({$\sim62\%$}) show LyC offsets. We note that the offset cut value of 0.1\arcsec{} can only give a lower limit fraction, as closer offsets can not be resolved especially with ground-based observations. 

In sum, $\sim$$60\%${--}$70\%$ of candidate LyC emitters at $z>3$ show spatial offsets of LyC according to current observations ({18/29} for the detections with HST resolution and {61/89} for all detections).

\subsection{Offset LyC in CDFS-6664}
\label{sec:offset_cdfs6664}

CDFS-6664 contains two components, C1 and C2, both with disk-like shapes and sizes comparable to galaxies at redshift 4, indicating that they are more likely to be galaxies than clumpy structures embedded in a single galaxy. With a projected separation of 0.92 kpc and a small radial velocity difference of about $195$ km/s, C1 and C2 are very likely in a merging process. The mass ratio of C1 to C2 (about 3.5) indicates a major merger. Both galaxies have sSFRs that exceed the average for star-forming galaxies at redshift 4. 

Our results show that the off-center LyC emission most likely originates from one of the components, C2, in CDFS-6664. SED fitting results show that C2 has younger stellar populations and a bluer UV slope than C1. All these properties of C2 are consistent with a high escape fraction of LyC photons. 

{The intense star-forming activities induced by merging can create an environment favoring the production and escape of ionizing photons \citep[e.g.][]{rauch2011,gupta2024}.} The outflows and clumpy interstellar medium common in mergers can cause the offset between the LyC and non-ionizing emission centers. The different escape fractions between the two galaxies further increase the observed offset, especially if the system is not spatially resolved. Compared to scenarios such as the substructures or infalling minor components, the merging scenario can apply to larger distance offsets since galaxy mergers can influence a larger scale than a single galaxy. 

The case of CDFS-6664 shows that galaxies with high escape fractions can exist in a merging system, with the LyC emission offset from the system center, building a connection between the merging process and the ionizing photon escape.  As increasing studies have shown that mergers are a more common process at higher redshifts \citep{duncan2019,asada2023,hsiao2023,gupta2023,witten2024}, it is reasonable to consider merging-induced LyC leaking as an important contributor to ionizing photons in the EoR. The interaction between two galaxies may enhance the star-forming activities and create a clumpy interstellar medium, allowing a large amount of LyC photons to escape from the system. The forthcoming JWST will provide more data on structures and properties of LyC emitters and unravel the questions of whether and to what degree mergers can contribute to the ionizing budget in the EoR.

\begin{acknowledgments}
{We thank the anonymous referees for useful comments and suggestions that helped improve the quality of this paper.} This work is supported by National Key R\&D Program of China No.2022YFF0503402. This work is partly supported by the Funds for Key Programs of Shanghai Astronomical Observatory. FTY acknowledges support from the Natural Science Foundation of Shanghai (Project Number: 21ZR1474300). ZYZ acknowledges support by the National Science Foundation of China (12022303), the China-Chile Joint Research Fund (CCJRF No. 1906). We also acknowledge the science research grants from the China Manned Space Project with NO. CMS-CSST-2021-A04, CMS-CSST-2021-A07. 

This work is based [in part] on observations made with the NASA/ESA/CSA James Webb Space Telescope. The data were obtained from the Mikulski Archive for Space Telescopes at the Space Telescope Science Institute, which is operated by the Association of Universities for Research in Astronomy, Inc., under NASA contract NAS 5-03127 for JWST. These observations are associated with program GO 1963 (PI: C. Williams). The authors acknowledge the JWST team of GO 1963 for developing their observing program with a zero-exclusive-access period, and Gabriel Brammer for making his reduction of the data available to the community.

This work is based [in part] on observations taken by the 3D-HST Treasury Program (GO 12177 and 12328) with the NASA/ESA HST, which is operated by the Association of Universities for Research in Astronomy, Inc., under NASA contract NAS5-26555.

All the {\it HST} data used in this paper can be found in MAST: \dataset[10.17909/T9JW9Z]{http://dx.doi.org/10.17909/T9JW9Z} \citep{https://doi.org/10.17909/t9jw9z} and \dataset[10.17909/T91019]{http://dx.doi.org/10.17909/T91019} \citep{https://doi.org/10.17909/t91019}.

\end{acknowledgments}

\appendix

\section{Appendix information}
\newpage

\startlongtable

\begin{deluxetable*}{>{\setlength{\baselineskip}{0.8\baselineskip}}m{0.5cm}<{\centering}
>{\setlength{\baselineskip}{0.8\baselineskip}}m{2.5cm}<{\centering}
>{\setlength{\baselineskip}{0.5\baselineskip}}m{3.8cm}<{\centering}
>{\setlength{\baselineskip}{0.8\baselineskip}}m{2.5cm}<{\centering}
>{\setlength{\baselineskip}{0.8\baselineskip}}m{2.5cm}<{\centering}
>{\setlength{\baselineskip}{0.8\baselineskip}}m{2.5cm}<{\centering}
>{\setlength{\baselineskip}{0.8\baselineskip}}m{0.5cm}<{\centering}
}

\tabletypesize{\scriptsize}
\tablecaption{Collection of candidate LyC emitters at $z_\mathrm{spec}>3$ in the literature. An offset is detected (marked as `Y') when the positional difference of the centers of the LyC and reference band {exceeds the maximum of 0.1\arcsec and the astrometric error.}}

\tablenum{A1}
\label{tab:lyc_cands}
\tablehead{\colhead{No.} & \colhead{ID} & \colhead{Ref.} & \colhead{$z_{spec}$} & \colhead{LyC band} & \colhead{Ref. band} & \colhead{Offset} \\
\colhead{(1)} & \colhead{(2)} & \colhead{(3)} & \colhead{(4)} & \colhead{(5)} & \colhead{(6)} & \colhead{(7)}
} 

\startdata
1 &   ion2 & \citet{vanzella2016}   & 3.212 &   HST F336W   &   HST F606W   &   N \\
2 &   ion1 & \citet{vanzella2012,ji2020} & 3.794 &   HST F410M   &   HST F775W   &   Y \\
3 &   CDFS-6664 & \citet{yuan2021}   & 3.797 &   HST F336W   &   HST F606W   &   Y \\
4 &   J0121+0025 & \citet{marques-chaves2021}   & 3.244 &   Spec\tnote{a}   &   Subaru SC/R   &   - \\
5 &   Q1549-C25 & \citet{shapley2016}   & 3.1526 &   Spec   &   HST F606W   &   - \\
6 &   ion3 & \citet{vanzella2018}   & 4 &   Spec   &   VLT HAWKI/Ks   &   - \\
7 &   CANDELS-5161 & \citet{saxena2022}   & 3.42 &      La Silla/NB396     &   HST F606W   &   Y \\
8 &   CANDELS-9358 & \citet{saxena2022}   & 3.229 &     CTIO/NB3727   &   HST F606W   &   - \\
9 &   CANDELS-9692 & \citet{saxena2022}   & 3.47 &      La Silla/NB396   &   HST F606W   &   Y \\
10 &   CANDELS-15718 & \citet{saxena2022}   & 3.439 &   La Silla/NB396   &   HST F606W   &   N \\
11 &   CANDELS-16444 & \citet{saxena2022}   & 3.128 &   CTIO/NB3727   &   HST F606W   &   N \\
12 &   CANDELS-18454 & \citet{saxena2022}   & 3.163 &   CTIO/NB3727   &   HST F606W   &   - \\
13 &   CANDELS-19872 & \citet{saxena2022}   & 3.452 &   La Silla/NB396   &   HST F606W   &   - \\
14 &   CANDELS-20745 & \citet{saxena2022}   & 3.495 &   La Silla/NB396   &   HST F606W   &   N \\
15 &   CANDELS-24975 & \citet{saxena2022}   & 3.187 &   CTIO/NB3727   &   HST F606W   &   - \\
16 &   F336W-189 & \citet{saxena2022,thorsen2022} & 3.46 &   HST F336W   &   HST F435W   &   Y \\
17 &   F336W–1041 & \citet{thorsen2022}   & 3.329 &   HST F336W   &   HST F435W   &   N \\
18 &   zf\_9775 & \citet{prichard2022}   & 4.365 &   HST F336W   &   HST F814W   &   Y \\
19 &   zf\_11754 & \citet{prichard2022}   & 4.47 &   HST F435W   &   HST F814W   &   Y \\
20 &   zf\_14000 & \citet{prichard2022}   & 3.727 &   HST F336W   &   HST F814W   &   N \\
21 & 1 & \citet{mestric2020}   & 4.28 &   CLAUDS(CFHT u)   &   HST F814W   &   N \\
22 & 326 & \citet{mestric2020}   & 3.57 &   CLAUDS(CFHT u)   &   HST F814W   &   N \\
23 & 330 & \citet{mestric2020}   & 5.09 &   CLAUDS(CFHT u)   &   HST F814W   &   N \\
24 & 368 & \citet{mestric2020}   & 3.64 &   CLAUDS(CFHT u)   &   HST F814W   &   Y \\
25 & 421 & \citet{mestric2020}   & 3.6 &   CLAUDS(CFHT u)   &   HST F814W   &   N \\
26 & 93564 & \citet{fletcher2019}   & 3.677 &   HST F336W   &   HST F160W   &   Y \\
27 & 90675 & \citet{fletcher2019}   & 3.111 &   HST F336W   &   HST F160W   &   N \\
28 & 92616 & \citet{fletcher2019}   & 3.0714 &   HST F336W   &   Subaru SC/NB497[Lya band]   &   Y \\
29 & 94460 & \citet{fletcher2019,nakajima2018, nakajima2020} & 3.0723 &   HST F336W   &   HST F160W   &   N \\
30 & 105937 & \citet{fletcher2019}   & 3.0668 &   HST F336W   &   HST F160W   &   Y \\
31 & 104037 & \citet{fletcher2019,nakajima2018, nakajima2020} & 3.065 &   HST F336W   &   HST F160W   &   Y \\
32 & 101846 & \citet{fletcher2019}   & 3.0565 &   HST F336W   &   HST F160W   &   Y \\
33 &   LAE2 & \citet{iwata2009}   &   3.07 $< z<$ 3.10   &   Subaru SC/NB359   &   HST F814W   &   Y \\
34 &   LAE3 & \citet{iwata2009}   &   3.07 $< z<$ 3.10   &   Subaru SC/NB359   &   HST F814W   &   Y \\
35 &   LAE5 & \citet{iwata2009}   &   3.07 $< z<$ 3.10   &   Subaru SC/NB359   &   Subaru SC/R   &   Y \\
36 &   LAE6 & \citet{iwata2009}   &   3.07 $< z<$ 3.10   &   Subaru SC/NB359   &   Subaru SC/R   &   Y \\
37 &   LAE7 & \citet{iwata2009}   &   3.07 $< z<$ 3.10   &   Subaru SC/NB359   &   Subaru SC/R   &   Y \\
38 &   LAE8 & \citet{iwata2009}   &   3.07 $< z<$ 3.10   &   Subaru SC/NB359   &   Subaru SC/R   &   Y \\
39 &   LAE9 & \citet{iwata2009}   &   3.07 $< z<$ 3.10   &   Subaru SC/NB359   &   Subaru SC/R   &   Y \\
40 &   LAE10 & \citet{iwata2009}   &   3.07 $< z<$ 3.10   &   Subaru SC/NB359   &   Subaru SC/R   &   Y \\
41 &   LBG3 & \citet{iwata2009}   &   3.04 $< z<$ 3.31   &   Subaru SC/NB359   &   HST F814W   &   Y \\
42 &   LBG4 & \citet{iwata2009}   &   3.04 $< z<$ 3.31   &   Subaru SC/NB359   &   HST F814W   &   Y \\
43 &   LBG5 & \citet{iwata2009}   &   3.04 $< z<$ 3.31   &   Subaru SC/NB359   &   HST F814W   &   Y \\
44 &   LBG6 & \citet{iwata2009}   &   3.04 $< z<$ 3.31   &   Subaru SC/NB359   &   Subaru SC/R   &   Y \\
45 &   R06-D23 & \citet{iwata2019}   & 3.123 &   Subaru SC/NB359/HST F336W   &   HST F606W   &   N \\
46 &   MOSDEF06336 & \citet{iwata2019}   & 3.412 &   Subaru SC/NB359/HST F336W   &   HST F606W   &   N \\
47 &   Q0933-M23 & \citet{steidel2018}   & 3.289 &   Spec   &   -   &   - \\
48 &   Westphal-CC38 & \citet{steidel2018}   & 3.0729 &   Spec   &   -   &   - \\
49 &   Westphal-MM37 & \citet{steidel2018}   & 3.4215 &   Spec   &   -   &   - \\
50 &   Q1422-d42 & \citet{steidel2018}   & 3.1369 &   Spec   &   -   &   - \\
51 &   Q1422-d68 & \citet{steidel2018}   & 3.2865 &   Spec   &   -   &   - \\
52 &  Q1549-C25 & \citet{steidel2018, shapley2016} & 3.1526 &   Spec   &   -   &   - \\
53 &   DSF2237b-MD38 & \citet{steidel2018}   & 3.3278 &   Spec   &   -   &   - \\
54 &   DSF2237b-MD60 & \citet{steidel2018}   & 3.1413 &   Spec   &   -   &   - \\
55 &   MD5b & \citet{mostardi2015}   & 3.143 &   HST F336W   &   HST F606W   &   Y \\
56 &   Micheva-LAE06 & \citet{micheva2017,inoue2011} & 3.075 &   Subaru SC/NB359   &   Subaru SC/R   &   Y \\
57 &   Micheva-LAE14 & \citet{micheva2017,inoue2011} & 3.095 &   Subaru SC/NB359   &   Subaru SC/R   &   Y \\
58 &   Micheva-LAE15 & \citet{micheva2017,inoue2011} & 3.094 &   Subaru SC/NB359   &   Subaru SC/R   &   Y \\
59 &   Micheva-LAE16 & \citet{micheva2017}   & 3.096 &   Subaru SC/NB359   &   Subaru SC/R   &   Y \\
60 &   Micheva-LAE17 & \citet{micheva2017}   & 3.089 &   Subaru SC/NB359   &   Subaru SC/R   &   Y \\
61 &   Micheva-LBG01 & \citet{micheva2017}   & 3.68 &   Subaru SC/NB359   &   Subaru SC/R   &   Y \\
62 &   Micheva-LBG02 & \citet{micheva2017}   & 3.113 &   Subaru SC/NB359   &   Subaru SC/R   &   Y \\
63 &   Micheva-LBG03 & \citet{micheva2017}   & 3.287 &   Subaru SC/NB359   &   Subaru SC/R   &   Y \\
64 &   Micheva-LBG04 & \citet{micheva2017}   & 3.311 &   Subaru SC/NB359   &   Subaru SC/R   &   Y  \\
65 &  LBG-C14 & \citet{nestor2013}  & 3.22 &  {Keck/NB3640}  &  Subaru SC/R  &  N \\
66 &  LBG-C16 & \citet{nestor2013}  & 3.065 &  {Keck/NB3640}  &  Subaru SC/R  &  Y \\
67 &  LBG-D17 & \citet{nestor2013, siana2015} & 3.09 &  {Keck/NB3640}  &  Subaru SC/R  &  Y \\
68 &  LBG-M2 & \citet{nestor2013}  & 3.388 &  {Keck/NB3640}  &  Subaru SC/R  &  Y \\
69 &  LBG-M29 & \citet{nestor2013}  & 3.228 &  {Keck/NB3640}  &  Subaru SC/R  &  Y \\
70 &  LAE010 & \citet{nestor2013}  & 3.096 &  {Keck/NB3640}  &  Subaru SC/R  &  Y \\
71 &  LAE016 & \citet{nestor2013}  & 3.091 &  {Keck/NB3640}  &  Subaru SC/R  &  Y \\
72 &  LAE018 & \citet{nestor2013, inoue2011, iwata2009} & 3.093 &  {Keck/NB3640}  &  Subaru SC/R  &  N \\
73 &  LAE028 & \citet{nestor2013}  & 3.088 &  {Keck/NB3640}  &  Subaru SC/R  &  Y \\
74 &  LAE039 & \citet{nestor2013}  & 3.095 &  {Keck/NB3640}  &  Subaru SC/R  &  Y \\
75 &  LAE041 & \citet{nestor2013}  & 3.066 & {Keck/NB3640}  &  Subaru SC/R  &  Y \\
76 &  LAE046 & \citet{nestor2013}  & 3.1 &  {Keck/NB3640}  &  Subaru SC/R  &  Y \\
77 &  LAE048 & \citet{nestor2013}  & 3.094 &  {Keck/NB3640}  &  Subaru SC/R  &  Y \\
78 &  LAE051 & \citet{nestor2013}  & 3.094 &  {Keck/NB3640}  &  Subaru SC/R  &  Y \\
79 &  LAE053 & \citet{nestor2013}  & 3.09 &  {Keck/NB3640}  &  Subaru SC/R  &  Y \\
80 &  LAE064 & \citet{nestor2013}  & 3.108 &  {Keck/NB3640}  &  Subaru SC/R  &  N \\
81 &  LAE069 & \citet{nestor2013}  & 3.074 &  {Keck/NB3640}  &  Subaru SC/R  &  Y \\
82 &  LAE074 & \citet{nestor2013}  & 3.105 &  {Keck/NB3640}  &  Subaru SC/R  &  N \\
83 &  LAE081 & \citet{nestor2013}  & 3.104 &  {Keck/NB3640}  &  Subaru SC/R  &  Y \\
84 & z19863  & \citet{gupta2024}   & 3.088 &  HST F336W
&  HST F160W    &  Y \\
85 & 1181371 & \citet{kerutt2024}  & 3.08  &  HST F336W
&  HST F160W    &  Y \\
86 & 109004028 & \citet{kerutt2024}  & 3.27  &  HST F336W
&  HST F160W    &  Y \\
87 & 126049137 & \citet{kerutt2024}  & 4.43  &  HST F336W
&  HST F160W    &  N \\
88 & 1521589 & \citet{kerutt2024}  & 3.15  &  HST F336W
&  HST F160W    &  Y \\
89 & 3452147 & \citet{kerutt2024}  & 3.52  &  HST F336W
&  HST F160W    &  Y \\
90 & 4062373 & \citet{kerutt2024}  & 3.66  &  HST F336W
&  HST F160W    &  Y \\
91 & 4172404 & \citet{kerutt2024}  & 3.67  &  HST F336W
&  HST F160W    &  Y \\
92 & 5622786 & \citet{kerutt2024}  & 4.00  &  HST F336W
&  HST F160W    &  N \\
93 & 119004004 & \citet{kerutt2024}  & 3.31  &  HST F336W
&  HST F160W    &  Y \\
94 & 122032127 & \citet{kerutt2024}  & 4.35  &  HST F336W
&  HST F160W    &  N \\
95 & LAE1      & \citet{liu2023}     & 3.167 &  Bok $U_J$
&  Subaru V    &  Y \\
96 & LAE2      & \citet{liu2023}     & 3.081 &  Bok $U_J$
&  Subaru V    &  N \\
97 & LAE3      & \citet{liu2023}     & 3.103 &  Bok $U_J$
&  Subaru V    &  Y \\
98 & LAE4      & \citet{liu2023}     & 3.120 &  Bok $U_J$
&  Subaru V    &  N \\
99 & LAE5      & \citet{liu2023}     & 3.065 &  Bok $U_J$
&  Subaru V    &  Y \\
{100} & {J1316+2614} & {\cite{marques-chaves2022,marques-chaves2024}} & {3.613} & {HST F410M} & {HST F775W} & {N}
\enddata
\tablecomments{(1) Number of LyC candidates; (2) ID from the references where the LyC candidate was detected; (3) References; (4) Spectroscopic redshift; (5) The band where LyC is detected. 'Spec' means LyC was detected by spectroscopy. (6) The reference band where the galaxy center is defined; (7) Offset: `Y' means there is an offset between the LyC and reference bands. `N' means no offset. `-' means we are unable to identify the offset due to insufficient data.}

\end{deluxetable*}

\bibliography{sample631}{}

\begin{thebibliography}{}
\expandafter\ifx\csname natexlab\endcsname\relax\def\natexlab#1{#1}\fi
\providecommand{\url}[1]{\href{#1}{#1}}
\providecommand{\dodoi}[1]{doi:~\href{http://doi.org/#1}{\nolinkurl{#1}}}
\providecommand{\doeprint}[1]{\href{http://ascl.net/#1}{\nolinkurl{http://ascl.net/#1}}}
\providecommand{\doarXiv}[1]{\href{https://arxiv.org/abs/#1}{\nolinkurl{https://arxiv.org/abs/#1}}}

\bibitem[{{Asada} {et~al.}(2023){Asada}, {Sawicki}, {Desprez}, {Abraham}, {Brada{\v{c}}}, {Brammer}, {Harshan}, {Iyer}, {Martis}, {Mowla}, {Muzzin}, {Noirot}, {Ravindranath}, {Sarrouh}, {Strait}, {Willott}, \& {Zabl}}]{asada2023}
{Asada}, Y., {Sawicki}, M., {Desprez}, G., {et~al.} 2023, \mnras, 523, L40, \dodoi{10.1093/mnrasl/slad054}

\bibitem[{{Bacon} {et~al.}(2010){Bacon}, {Accardo}, {Adjali}, {Anwand}, {Bauer}, {Biswas}, {Blaizot}, {Boudon}, {Brau-Nogue}, {Brinchmann}, {Caillier}, {Capoani}, {Carollo}, {Contini}, {Couderc}, {Daguis{\'e}}, {Deiries}, {Delabre}, {Dreizler}, {Dubois}, {Dupieux}, {Dupuy}, {Emsellem}, {Fechner}, {Fleischmann}, {Fran{\c{c}}ois}, {Gallou}, {Gharsa}, {Glindemann}, {Gojak}, {Guiderdoni}, {Hansali}, {Hahn}, {Jarno}, {Kelz}, {Koehler}, {Kosmalski}, {Laurent}, {Le Floch}, {Lilly}, {Lizon}, {Loupias}, {Manescau}, {Monstein}, {Nicklas}, {Olaya}, {Pares}, {Pasquini}, {P{\'e}contal-Rousset}, {Pell{\'o}}, {Petit}, {Popow}, {Reiss}, {Remillieux}, {Renault}, {Roth}, {Rupprecht}, {Serre}, {Schaye}, {Soucail}, {Steinmetz}, {Streicher}, {Stuik}, {Valentin}, {Vernet}, {Weilbacher}, {Wisotzki}, \& {Yerle}}]{bacon2010}
{Bacon}, R., {Accardo}, M., {Adjali}, L., {et~al.} 2010, in Society of Photo-Optical Instrumentation Engineers (SPIE) Conference Series, Vol. 7735, Ground-based and Airborne Instrumentation for Astronomy III, ed. I.~S. {McLean}, S.~K. {Ramsay}, \& H.~{Takami}, 773508, \dodoi{10.1117/12.856027}

\bibitem[{{Bera} {et~al.}(2023){Bera}, {Hassan}, {Smith}, {Cen}, {Garaldi}, {Kannan}, \& {Vogelsberger}}]{bera2023}
{Bera}, A., {Hassan}, S., {Smith}, A., {et~al.} 2023, \apj, 959, 2, \dodoi{10.3847/1538-4357/ad05c0}

\bibitem[{{Boquien} {et~al.}(2019){Boquien}, {Burgarella}, {Roehlly}, {Buat}, {Ciesla}, {Corre}, {Inoue}, \& {Salas}}]{boquien2019}
{Boquien}, M., {Burgarella}, D., {Roehlly}, Y., {et~al.} 2019, \aap, 622, A103, \dodoi{10.1051/0004-6361/201834156}

\bibitem[{{Bruzual} \& {Charlot}(2003)}]{bc2003}
{Bruzual}, G., \& {Charlot}, S. 2003, \mnras, 344, 1000, \dodoi{10.1046/j.1365-8711.2003.06897.x}

\bibitem[{{Burgarella} {et~al.}(2005){Burgarella}, {Buat}, \& {Iglesias-P{\'a}ramo}}]{burgarella2005}
{Burgarella}, D., {Buat}, V., \& {Iglesias-P{\'a}ramo}, J. 2005, \mnras, 360, 1413, \dodoi{10.1111/j.1365-2966.2005.09131.x}

\bibitem[{{Calzetti} {et~al.}(2000){Calzetti}, {Armus}, {Bohlin}, {Kinney}, {Koornneef}, \& {Storchi-Bergmann}}]{calzetti2000}
{Calzetti}, D., {Armus}, L., {Bohlin}, R.~C., {et~al.} 2000, \apj, 533, 682, \dodoi{10.1086/308692}

\bibitem[{{Cardelli} {et~al.}(1989){Cardelli}, {Clayton}, \& {Mathis}}]{cardelli1989}
{Cardelli}, J.~A., {Clayton}, G.~C., \& {Mathis}, J.~S. 1989, \apj, 345, 245, \dodoi{10.1086/167900}

\bibitem[{{Chisholm} {et~al.}(2022){Chisholm}, {Saldana-Lopez}, {Flury}, {Schaerer}, {Jaskot}, {Amor{\'\i}n}, {Atek}, {Finkelstein}, {Fleming}, {Ferguson}, {Fern{\'a}ndez}, {Giavalisco}, {Hayes}, {Heckman}, {Henry}, {Ji}, {Marques-Chaves}, {Mauerhofer}, {McCandliss}, {Oey}, {{\"O}stlin}, {Rutkowski}, {Scarlata}, {Thuan}, {Trebitsch}, {Wang}, {Worseck}, \& {Xu}}]{chisholm2022}
{Chisholm}, J., {Saldana-Lopez}, A., {Flury}, S., {et~al.} 2022, \mnras, 517, 5104, \dodoi{10.1093/mnras/stac2874}

\bibitem[{{Davidzon} {et~al.}(2017){Davidzon}, {Ilbert}, {Laigle}, {Coupon}, {McCracken}, {Delvecchio}, {Masters}, {Capak}, {Hsieh}, {Le F{\`e}vre}, {Tresse}, {Bethermin}, {Chang}, {Faisst}, {Le Floc'h}, {Steinhardt}, {Toft}, {Aussel}, {Dubois}, {Hasinger}, {Salvato}, {Sanders}, {Scoville}, \& {Silverman}}]{davidzon2017}
{Davidzon}, I., {Ilbert}, O., {Laigle}, C., {et~al.} 2017, \aap, 605, A70, \dodoi{10.1051/0004-6361/201730419}

\bibitem[{{de Barros} {et~al.}(2014){de Barros}, {Schaerer}, \& {Stark}}]{debarros2014}
{de Barros}, S., {Schaerer}, D., \& {Stark}, D.~P. 2014, \aap, 563, A81, \dodoi{10.1051/0004-6361/201220026}

\bibitem[{{Duncan} {et~al.}(2019){Duncan}, {Conselice}, {Mundy}, {Bell}, {Donley}, {Galametz}, {Guo}, {Grogin}, {Hathi}, {Kartaltepe}, {Kocevski}, {Koekemoer}, {P{\'e}rez-Gonz{\'a}lez}, {Mantha}, {Snyder}, \& {Stefanon}}]{duncan2019}
{Duncan}, K., {Conselice}, C.~J., {Mundy}, C., {et~al.} 2019, \apj, 876, 110, \dodoi{10.3847/1538-4357/ab148a}

\bibitem[{{Eisenhauer} {et~al.}(2003){Eisenhauer}, {Abuter}, {Bickert}, {Biancat-Marchet}, {Bonnet}, {Brynnel}, {Conzelmann}, {Delabre}, {Donaldson}, {Farinato}, {Fedrigo}, {Genzel}, {Hubin}, {Iserlohe}, {Kasper}, {Kissler-Patig}, {Monnet}, {Roehrle}, {Schreiber}, {Stroebele}, {Tecza}, {Thatte}, \& {Weisz}}]{eisenhauer2003}
{Eisenhauer}, F., {Abuter}, R., {Bickert}, K., {et~al.} 2003, in Society of Photo-Optical Instrumentation Engineers (SPIE) Conference Series, Vol. 4841, Instrument Design and Performance for Optical/Infrared Ground-based Telescopes, ed. M.~{Iye} \& A.~F.~M. {Moorwood}, 1548--1561, \dodoi{10.1117/12.459468}

\bibitem[{{Erb}(2015)}]{erb2015}
{Erb}, D.~K. 2015, \nat, 523, 169, \dodoi{10.1038/nature14454}

\bibitem[{{Finkelstein} {et~al.}(2012){Finkelstein}, {Papovich}, {Ryan}, {Pawlik}, {Dickinson}, {Ferguson}, {Finlator}, {Koekemoer}, {Giavalisco}, {Cooray}, {Dunlop}, {Faber}, {Grogin}, {Kocevski}, \& {Newman}}]{finkelstein2012}
{Finkelstein}, S.~L., {Papovich}, C., {Ryan}, R.~E., {et~al.} 2012, \apj, 758, 93, \dodoi{10.1088/0004-637X/758/2/93}

\bibitem[{{Finkelstein} {et~al.}(2019){Finkelstein}, {D'Aloisio}, {Paardekooper}, {Ryan}, {Behroozi}, {Finlator}, {Livermore}, {Upton Sanderbeck}, {Dalla Vecchia}, \& {Khochfar}}]{finkelstein2019}
{Finkelstein}, S.~L., {D'Aloisio}, A., {Paardekooper}, J.-P., {et~al.} 2019, \apj, 879, 36, \dodoi{10.3847/1538-4357/ab1ea8}

\bibitem[{{Fletcher} {et~al.}(2019){Fletcher}, {Tang}, {Robertson}, {Nakajima}, {Ellis}, {Stark}, \& {Inoue}}]{fletcher2019}
{Fletcher}, T.~J., {Tang}, M., {Robertson}, B.~E., {et~al.} 2019, \apj, 878, 87, \dodoi{10.3847/1538-4357/ab2045}

\bibitem[{{Gnedin} {et~al.}(2008){Gnedin}, {Kravtsov}, \& {Chen}}]{gnedin2008}
{Gnedin}, N.~Y., {Kravtsov}, A.~V., \& {Chen}, H.-W. 2008, \apj, 672, 765, \dodoi{10.1086/524007}

\bibitem[{{Gnerucci} {et~al.}(2011){Gnerucci}, {Marconi}, {Cresci}, {Maiolino}, {Mannucci}, {Calura}, {Cimatti}, {Cocchia}, {Grazian}, {Matteucci}, {Nagao}, {Pozzetti}, \& {Troncoso}}]{gnerucci2011}
{Gnerucci}, A., {Marconi}, A., {Cresci}, G., {et~al.} 2011, \aap, 528, A88, \dodoi{10.1051/0004-6361/201015465}

\bibitem[{{Grogin} {et~al.}(2011){Grogin}, {Kocevski}, {Faber}, {Ferguson}, {Koekemoer}, {Riess}, {Acquaviva}, {Alexander}, {Almaini}, {Ashby}, {Barden}, {Bell}, {Bournaud}, {Brown}, {Caputi}, {Casertano}, {Cassata}, {Castellano}, {Challis}, {Chary}, {Cheung}, {Cirasuolo}, {Conselice}, {Roshan Cooray}, {Croton}, {Daddi}, {Dahlen}, {Dav{\'e}}, {de Mello}, {Dekel}, {Dickinson}, {Dolch}, {Donley}, {Dunlop}, {Dutton}, {Elbaz}, {Fazio}, {Filippenko}, {Finkelstein}, {Fontana}, {Gardner}, {Garnavich}, {Gawiser}, {Giavalisco}, {Grazian}, {Guo}, {Hathi}, {H{\"a}ussler}, {Hopkins}, {Huang}, {Huang}, {Jha}, {Kartaltepe}, {Kirshner}, {Koo}, {Lai}, {Lee}, {Li}, {Lotz}, {Lucas}, {Madau}, {McCarthy}, {McGrath}, {McIntosh}, {McLure}, {Mobasher}, {Moustakas}, {Mozena}, {Nandra}, {Newman}, {Niemi}, {Noeske}, {Papovich}, {Pentericci}, {Pope}, {Primack}, {Rajan}, {Ravindranath}, {Reddy}, {Renzini}, {Rix}, {Robaina}, {Rodney}, {Rosario}, {Rosati}, {Salimbeni}, {Scarlata}, {Siana}, {Simard}, {Smidt}, {Somerville}, {Spinrad},
  {Straughn}, {Strolger}, {Telford}, {Teplitz}, {Trump}, {van der Wel}, {Villforth}, {Wechsler}, {Weiner}, {Wiklind}, {Wild}, {Wilson}, {Wuyts}, {Yan}, \& {Yun}}]{grogin2011}
{Grogin}, N.~A., {Kocevski}, D.~D., {Faber}, S.~M., {et~al.} 2011, \apjs, 197, 35, \dodoi{10.1088/0067-0049/197/2/35}

\bibitem[{{Guo} {et~al.}(2012){Guo}, {Giavalisco}, {Ferguson}, {Cassata}, \& {Koekemoer}}]{guo2012}
{Guo}, Y., {Giavalisco}, M., {Ferguson}, H.~C., {Cassata}, P., \& {Koekemoer}, A.~M. 2012, \apj, 757, 120, \dodoi{10.1088/0004-637X/757/2/120}

\bibitem[{{Gupta} {et~al.}(2023){Gupta}, {Tran}, {Mendel}, {Harshan}, {Forrest}, {Davies}, {Wisnioski}, {Nanayakkara}, {Kacprzak}, \& {Kewley}}]{gupta2023}
{Gupta}, A., {Tran}, K.-V., {Mendel}, T., {et~al.} 2023, \mnras, 519, 980, \dodoi{10.1093/mnras/stac3548}

\bibitem[{{Gupta} {et~al.}(2024){Gupta}, {Trott}, {Jaiswar}, {Ryan-Weber}, {Bunker}, {Acharyya}, {Cameron}, {Forrest}, {Kacprzak}, {Nanayakkara}, {Tran}, \& {Chokshi}}]{gupta2024}
{Gupta}, A., {Trott}, C.~M., {Jaiswar}, R., {et~al.} 2024, arXiv e-prints, arXiv:2403.13285, \dodoi{10.48550/arXiv.2403.13285}

\bibitem[{{Hsiao} {et~al.}(2023){Hsiao}, {Coe}, {Abdurro'uf}, {Whitler}, {Jung}, {Khullar}, {Meena}, {Dayal}, {Barrow}, {Santos-Olmsted}, {Casselman}, {Vanzella}, {Nonino}, {Jim{\'e}nez-Teja}, {Oguri}, {Stark}, {Furtak}, {Zitrin}, {Adamo}, {Brammer}, {Bradley}, {Diego}, {Zackrisson}, {Finkelstein}, {Windhorst}, {Bhatawdekar}, {Hutchison}, {Broadhurst}, {Dimauro}, {Andrade-Santos}, {Eldridge}, {Acebron}, {Avila}, {Bayliss}, {Ben{\'\i}tez}, {Binggeli}, {Bolan}, {Brada{\v{c}}}, {Carnall}, {Conselice}, {Donahue}, {Frye}, {Fujimoto}, {Henry}, {James}, {Kassin}, {Kewley}, {Larson}, {Lauer}, {Law}, {Mahler}, {Mainali}, {McCandliss}, {Nicholls}, {Pirzkal}, {Postman}, {Rigby}, {Ryan}, {Senchyna}, {Sharon}, {Shimizu}, {Strait}, {Tang}, {Trenti}, {Vikaeus}, \& {Welch}}]{hsiao2023}
{Hsiao}, T. Y.-Y., {Coe}, D., {Abdurro'uf}, {et~al.} 2023, \apjl, 949, L34, \dodoi{10.3847/2041-8213/acc94b}

\bibitem[{Illingworth(2015)}]{https://doi.org/10.17909/t91019}
Illingworth, G. 2015, Hubble Legacy Fields ("HLF"),  STScI/MAST, \dodoi{10.17909/T91019}

\bibitem[{{Illingworth} {et~al.}(2016){Illingworth}, {Magee}, {Bouwens}, {Oesch}, {Labbe}, {van Dokkum}, {Whitaker}, {Holden}, {Franx}, \& {Gonzalez}}]{illingworth2016}
{Illingworth}, G., {Magee}, D., {Bouwens}, R., {et~al.} 2016, arXiv e-prints, arXiv:1606.00841, \dodoi{10.48550/arXiv.1606.00841}

\bibitem[{{Inoue}(2010)}]{inoue2010}
{Inoue}, A.~K. 2010, \mnras, 401, 1325, \dodoi{10.1111/j.1365-2966.2009.15730.x}

\bibitem[{{Inoue}(2011)}]{inoue2011b}
---. 2011, \mnras, 415, 2920, \dodoi{10.1111/j.1365-2966.2011.18906.x}

\bibitem[{{Inoue} {et~al.}(2011){Inoue}, {Kousai}, {Iwata}, {Matsuda}, {Nakamura}, {Horie}, {Hayashino}, {Tapken}, {Akiyama}, {Noll}, {Yamada}, {Burgarella}, \& {Nakamura}}]{inoue2011}
{Inoue}, A.~K., {Kousai}, K., {Iwata}, I., {et~al.} 2011, \mnras, 411, 2336, \dodoi{10.1111/j.1365-2966.2010.17851.x}

\bibitem[{{Iwata} {et~al.}(2019){Iwata}, {Inoue}, {Micheva}, {Matsuda}, \& {Yamada}}]{iwata2019}
{Iwata}, I., {Inoue}, A.~K., {Micheva}, G., {Matsuda}, Y., \& {Yamada}, T. 2019, \mnras, 488, 5671, \dodoi{10.1093/mnras/stz2081}

\bibitem[{{Iwata} {et~al.}(2009){Iwata}, {Inoue}, {Matsuda}, {Furusawa}, {Hayashino}, {Kousai}, {Akiyama}, {Yamada}, {Burgarella}, \& {Deharveng}}]{iwata2009}
{Iwata}, I., {Inoue}, A.~K., {Matsuda}, Y., {et~al.} 2009, \apj, 692, 1287, \dodoi{10.1088/0004-637X/692/2/1287}

\bibitem[{{Ji} {et~al.}(2020){Ji}, {Giavalisco}, {Vanzella}, {Siana}, {Pentericci}, {Jaskot}, {Liu}, {Nonino}, {Ferguson}, {Castellano}, {Mannucci}, {Schaerer}, {Fynbo}, {Papovich}, {Carnall}, {Amorin}, {Simons}, {Hathi}, {Cullen}, \& {McLeod}}]{ji2020}
{Ji}, Z., {Giavalisco}, M., {Vanzella}, E., {et~al.} 2020, \apj, 888, 109, \dodoi{10.3847/1538-4357/ab5fdc}

\bibitem[{{Jiang} {et~al.}(2022){Jiang}, {Ning}, {Fan}, {Ho}, {Luo}, {Wang}, {Wu}, {Wu}, {Yang}, \& {Zheng}}]{jiang2022}
{Jiang}, L., {Ning}, Y., {Fan}, X., {et~al.} 2022, Nature Astronomy, 6, 850, \dodoi{10.1038/s41550-022-01708-w}

\bibitem[{{Kakiichi} \& {Gronke}(2021)}]{kakiichi2021}
{Kakiichi}, K., \& {Gronke}, M. 2021, \apj, 908, 30, \dodoi{10.3847/1538-4357/abc2d9}

\bibitem[{{Kerutt} {et~al.}(2024){Kerutt}, {Oesch}, {Wisotzki}, {Verhamme}, {Atek}, {Herenz}, {Illingworth}, {Kusakabe}, {Matthee}, {Mauerhofer}, {Montes}, {Naidu}, {Nelson}, {Reddy}, {Schaye}, {Simmonds}, {Urrutia}, \& {Vitte}}]{kerutt2024}
{Kerutt}, J., {Oesch}, P.~A., {Wisotzki}, L., {et~al.} 2024, \aap, 684, A42, \dodoi{10.1051/0004-6361/202346656}

\bibitem[{{Koekemoer} {et~al.}(2011){Koekemoer}, {Faber}, {Ferguson}, {Grogin}, {Kocevski}, {Koo}, {Lai}, {Lotz}, {Lucas}, {McGrath}, {Ogaz}, {Rajan}, {Riess}, {Rodney}, {Strolger}, {Casertano}, {Castellano}, {Dahlen}, {Dickinson}, {Dolch}, {Fontana}, {Giavalisco}, {Grazian}, {Guo}, {Hathi}, {Huang}, {van der Wel}, {Yan}, {Acquaviva}, {Alexander}, {Almaini}, {Ashby}, {Barden}, {Bell}, {Bournaud}, {Brown}, {Caputi}, {Cassata}, {Challis}, {Chary}, {Cheung}, {Cirasuolo}, {Conselice}, {Roshan Cooray}, {Croton}, {Daddi}, {Dav{\'e}}, {de Mello}, {de Ravel}, {Dekel}, {Donley}, {Dunlop}, {Dutton}, {Elbaz}, {Fazio}, {Filippenko}, {Finkelstein}, {Frazer}, {Gardner}, {Garnavich}, {Gawiser}, {Gruetzbauch}, {Hartley}, {H{\"a}ussler}, {Herrington}, {Hopkins}, {Huang}, {Jha}, {Johnson}, {Kartaltepe}, {Khostovan}, {Kirshner}, {Lani}, {Lee}, {Li}, {Madau}, {McCarthy}, {McIntosh}, {McLure}, {McPartland}, {Mobasher}, {Moreira}, {Mortlock}, {Moustakas}, {Mozena}, {Nandra}, {Newman}, {Nielsen}, {Niemi}, {Noeske}, {Papovich},
  {Pentericci}, {Pope}, {Primack}, {Ravindranath}, {Reddy}, {Renzini}, {Rix}, {Robaina}, {Rosario}, {Rosati}, {Salimbeni}, {Scarlata}, {Siana}, {Simard}, {Smidt}, {Snyder}, {Somerville}, {Spinrad}, {Straughn}, {Telford}, {Teplitz}, {Trump}, {Vargas}, {Villforth}, {Wagner}, {Wandro}, {Wechsler}, {Weiner}, {Wiklind}, {Wild}, {Wilson}, {Wuyts}, \& {Yun}}]{koekemoer2011}
{Koekemoer}, A.~M., {Faber}, S.~M., {Ferguson}, H.~C., {et~al.} 2011, \apjs, 197, 36, \dodoi{10.1088/0067-0049/197/2/36}

\bibitem[{{Lange} {et~al.}(2016){Lange}, {Moffett}, {Driver}, {Robotham}, {Lagos}, {Kelvin}, {Conselice}, {Margalef-Bentabol}, {Alpaslan}, {Baldry}, {Bland-Hawthorn}, {Bremer}, {Brough}, {Cluver}, {Colless}, {Davies}, {H{\"a}u{\ss}ler}, {Holwerda}, {Hopkins}, {Kafle}, {Kennedy}, {Liske}, {Phillipps}, {Popescu}, {Taylor}, {Tuffs}, {van Kampen}, \& {Wright}}]{lange2016}
{Lange}, R., {Moffett}, A.~J., {Driver}, S.~P., {et~al.} 2016, \mnras, 462, 1470, \dodoi{10.1093/mnras/stw1495}

\bibitem[{{Liu} {et~al.}(2023){Liu}, {Jiang}, {Windhorst}, {Guo}, \& {Zheng}}]{liu2023}
{Liu}, Y., {Jiang}, L., {Windhorst}, R.~A., {Guo}, Y., \& {Zheng}, Z.-Y. 2023, \apj, 958, 22, \dodoi{10.3847/1538-4357/acf9fa}

\bibitem[{{Madau} {et~al.}(1999){Madau}, {Haardt}, \& {Rees}}]{madau1999}
{Madau}, P., {Haardt}, F., \& {Rees}, M.~J. 1999, \apj, 514, 648, \dodoi{10.1086/306975}

\bibitem[{{Maiolino} {et~al.}(2008){Maiolino}, {Nagao}, {Grazian}, {Cocchia}, {Marconi}, {Mannucci}, {Cimatti}, {Pipino}, {Ballero}, {Calura}, {Chiappini}, {Fontana}, {Granato}, {Matteucci}, {Pastorini}, {Pentericci}, {Risaliti}, {Salvati}, \& {Silva}}]{maiolino2008}
{Maiolino}, R., {Nagao}, T., {Grazian}, A., {et~al.} 2008, \aap, 488, 463, \dodoi{10.1051/0004-6361:200809678}

\bibitem[{{Marques-Chaves} {et~al.}(2021){Marques-Chaves}, {Schaerer}, {{\'A}lvarez-M{\'a}rquez}, {Colina}, {Dessauges-Zavadsky}, {P{\'e}rez-Fournon}, {Saldana-Lopez}, \& {Verhamme}}]{marques-chaves2021}
{Marques-Chaves}, R., {Schaerer}, D., {{\'A}lvarez-M{\'a}rquez}, J., {et~al.} 2021, \mnras, 507, 524, \dodoi{10.1093/mnras/stab2187}

\bibitem[{{Marques-Chaves} {et~al.}(2022){Marques-Chaves}, {Schaerer}, {{\'A}lvarez-M{\'a}rquez}, {Verhamme}, {Ceverino}, {Chisholm}, {Colina}, {Dessauges-Zavadsky}, {P{\'e}rez-Fournon}, {Saldana-Lopez}, {Upadhyaya}, \& {Vanzella}}]{marques-chaves2022}
---. 2022, \mnras, 517, 2972, \dodoi{10.1093/mnras/stac2893}

\bibitem[{{Marques-Chaves} {et~al.}(2024){Marques-Chaves}, {Schaerer}, {Vanzella}, {Verhamme}, {Dessauges-Zavadsky}, {Chisholm}, {Leclercq}, {Upadhyaya}, {Alvarez-Marquez}, {Colina}, {Garel}, \& {Messa}}]{marques-chaves2024}
{Marques-Chaves}, R., {Schaerer}, D., {Vanzella}, E., {et~al.} 2024, arXiv e-prints, arXiv:2407.18804, \dodoi{10.48550/arXiv.2407.18804}

\bibitem[{{Matsuoka} {et~al.}(2018){Matsuoka}, {Strauss}, {Kashikawa}, {Onoue}, {Iwasawa}, {Tang}, {Lee}, {Imanishi}, {Nagao}, {Akiyama}, {Asami}, {Bosch}, {Furusawa}, {Goto}, {Gunn}, {Harikane}, {Ikeda}, {Izumi}, {Kawaguchi}, {Kato}, {Kikuta}, {Kohno}, {Komiyama}, {Lupton}, {Minezaki}, {Miyazaki}, {Murayama}, {Niida}, {Nishizawa}, {Noboriguchi}, {Oguri}, {Ono}, {Ouchi}, {Price}, {Sameshima}, {Schulze}, {Shirakata}, {Silverman}, {Sugiyama}, {Tait}, {Takada}, {Takata}, {Tanaka}, {Toba}, {Utsumi}, {Wang}, \& {Yamashita}}]{matsuoka2018}
{Matsuoka}, Y., {Strauss}, M.~A., {Kashikawa}, N., {et~al.} 2018, \apj, 869, 150, \dodoi{10.3847/1538-4357/aaee7a}

\bibitem[{{Matsuoka} {et~al.}(2023){Matsuoka}, {Onoue}, {Iwasawa}, {Strauss}, {Kashikawa}, {Izumi}, {Nagao}, {Imanishi}, {Akiyama}, {Silverman}, {Asami}, {Bosch}, {Furusawa}, {Goto}, {Gunn}, {Harikane}, {Ikeda}, {Inayoshi}, {Ishimoto}, {Kawaguchi}, {Kikuta}, {Kohno}, {Komiyama}, {Lee}, {Lupton}, {Minezaki}, {Miyazaki}, {Murayama}, {Nishizawa}, {Oguri}, {Ono}, {Oogi}, {Ouchi}, {Price}, {Sameshima}, {Sugiyama}, {Tait}, {Takada}, {Takahashi}, {Takata}, {Tanaka}, {Toba}, {Wang}, \& {Yamashita}}]{matsuoka2023}
{Matsuoka}, Y., {Onoue}, M., {Iwasawa}, K., {et~al.} 2023, \apjl, 949, L42, \dodoi{10.3847/2041-8213/acd69f}

\bibitem[{{Meiksin}(2006)}]{meiksin2006}
{Meiksin}, A. 2006, \mnras, 365, 807, \dodoi{10.1111/j.1365-2966.2005.09756.x}

\bibitem[{{Me{\v{s}}tri{\'c}} {et~al.}(2020){Me{\v{s}}tri{\'c}}, {Ryan-Weber}, {Cooke}, {Bassett}, {Sawicki}, {Faisst}, {Kakiichi}, {Inoue}, {Rafelski}, {Prichard}, {Arnouts}, {Moutard}, {Coupon}, {Golob}, \& {Gwyn}}]{mestric2020}
{Me{\v{s}}tri{\'c}}, U., {Ryan-Weber}, E.~V., {Cooke}, J., {et~al.} 2020, \mnras, 494, 4986, \dodoi{10.1093/mnras/staa920}

\bibitem[{{Micheva} {et~al.}(2017){Micheva}, {Iwata}, {Inoue}, {Matsuda}, {Yamada}, \& {Hayashino}}]{micheva2017}
{Micheva}, G., {Iwata}, I., {Inoue}, A.~K., {et~al.} 2017, \mnras, 465, 316, \dodoi{10.1093/mnras/stw2700}

\bibitem[{Momcheva(2017)}]{https://doi.org/10.17909/t9jw9z}
Momcheva, I. 2017, 3D-HST,  STScI/MAST, \dodoi{10.17909/T9JW9Z}

\bibitem[{{Mostardi} {et~al.}(2013){Mostardi}, {Shapley}, {Nestor}, {Steidel}, {Reddy}, \& {Trainor}}]{mostardi2013}
{Mostardi}, R.~E., {Shapley}, A.~E., {Nestor}, D.~B., {et~al.} 2013, \apj, 779, 65, \dodoi{10.1088/0004-637X/779/1/65}

\bibitem[{{Mostardi} {et~al.}(2015){Mostardi}, {Shapley}, {Steidel}, {Trainor}, {Reddy}, \& {Siana}}]{mostardi2015}
{Mostardi}, R.~E., {Shapley}, A.~E., {Steidel}, C.~C., {et~al.} 2015, \apj, 810, 107, \dodoi{10.1088/0004-637X/810/2/107}

\bibitem[{{Naidu} {et~al.}(2022){Naidu}, {Matthee}, {Oesch}, {Conroy}, {Sobral}, {Pezzulli}, {Hayes}, {Erb}, {Amor{\'\i}n}, {Gronke}, {Schaerer}, {Tacchella}, {Kerutt}, {Paulino-Afonso}, {Calhau}, {Llerena}, \& {R{\"o}ttgering}}]{naidu2022}
{Naidu}, R.~P., {Matthee}, J., {Oesch}, P.~A., {et~al.} 2022, \mnras, 510, 4582, \dodoi{10.1093/mnras/stab3601}

\bibitem[{{Nakajima} {et~al.}(2020){Nakajima}, {Ellis}, {Robertson}, {Tang}, \& {Stark}}]{nakajima2020}
{Nakajima}, K., {Ellis}, R.~S., {Robertson}, B.~E., {Tang}, M., \& {Stark}, D.~P. 2020, \apj, 889, 161, \dodoi{10.3847/1538-4357/ab6604}

\bibitem[{{Nakajima} {et~al.}(2018){Nakajima}, {Fletcher}, {Ellis}, {Robertson}, \& {Iwata}}]{nakajima2018}
{Nakajima}, K., {Fletcher}, T., {Ellis}, R.~S., {Robertson}, B.~E., \& {Iwata}, I. 2018, \mnras, 477, 2098, \dodoi{10.1093/mnras/sty750}

\bibitem[{{Nestor} {et~al.}(2013){Nestor}, {Shapley}, {Kornei}, {Steidel}, \& {Siana}}]{nestor2013}
{Nestor}, D.~B., {Shapley}, A.~E., {Kornei}, K.~A., {Steidel}, C.~C., \& {Siana}, B. 2013, \apj, 765, 47, \dodoi{10.1088/0004-637X/765/1/47}

\bibitem[{{Noll} {et~al.}(2009){Noll}, {Burgarella}, {Giovannoli}, {Buat}, {Marcillac}, \& {Mu{\~n}oz-Mateos}}]{noll2009}
{Noll}, S., {Burgarella}, D., {Giovannoli}, E., {et~al.} 2009, \aap, 507, 1793, \dodoi{10.1051/0004-6361/200912497}

\bibitem[{{Oesch} {et~al.}(2018){Oesch}, {Montes}, {Reddy}, {Bouwens}, {Illingworth}, {Magee}, {Atek}, {Carollo}, {Cibinel}, {Franx}, {Holden}, {Labb{\'e}}, {Nelson}, {Steidel}, {van Dokkum}, {Morselli}, {Naidu}, \& {Wilkins}}]{oesch2018}
{Oesch}, P.~A., {Montes}, M., {Reddy}, N., {et~al.} 2018, \apjs, 237, 12, \dodoi{10.3847/1538-4365/aacb30}

\bibitem[{{Ouchi} {et~al.}(2009){Ouchi}, {Mobasher}, {Shimasaku}, {Ferguson}, {Fall}, {Ono}, {Kashikawa}, {Morokuma}, {Nakajima}, {Okamura}, {Dickinson}, {Giavalisco}, \& {Ohta}}]{ouchi2009}
{Ouchi}, M., {Mobasher}, B., {Shimasaku}, K., {et~al.} 2009, \apj, 706, 1136, \dodoi{10.1088/0004-637X/706/2/1136}

\bibitem[{{Perrin} {et~al.}(2014){Perrin}, {Sivaramakrishnan}, {Lajoie}, {Elliott}, {Pueyo}, {Ravindranath}, \& {Albert}}]{perrin2014}
{Perrin}, M.~D., {Sivaramakrishnan}, A., {Lajoie}, C.-P., {et~al.} 2014, in Society of Photo-Optical Instrumentation Engineers (SPIE) Conference Series, Vol. 9143, Space Telescopes and Instrumentation 2014: Optical, Infrared, and Millimeter Wave, ed. J.~{Oschmann}, Jacobus~M., M.~{Clampin}, G.~G. {Fazio}, \& H.~A. {MacEwen}, 91433X, \dodoi{10.1117/12.2056689}

\bibitem[{{Perrin} {et~al.}(2012){Perrin}, {Soummer}, {Elliott}, {Lallo}, \& {Sivaramakrishnan}}]{perrin2012}
{Perrin}, M.~D., {Soummer}, R., {Elliott}, E.~M., {Lallo}, M.~D., \& {Sivaramakrishnan}, A. 2012, in Society of Photo-Optical Instrumentation Engineers (SPIE) Conference Series, Vol. 8442, Space Telescopes and Instrumentation 2012: Optical, Infrared, and Millimeter Wave, ed. M.~C. {Clampin}, G.~G. {Fazio}, H.~A. {MacEwen}, \& J.~{Oschmann}, Jacobus~M., 84423D, \dodoi{10.1117/12.925230}

\bibitem[{{Popesso} {et~al.}(2023){Popesso}, {Concas}, {Cresci}, {Belli}, {Rodighiero}, {Inami}, {Dickinson}, {Ilbert}, {Pannella}, \& {Elbaz}}]{popesso2023}
{Popesso}, P., {Concas}, A., {Cresci}, G., {et~al.} 2023, \mnras, 519, 1526, \dodoi{10.1093/mnras/stac3214}

\bibitem[{{Prichard} {et~al.}(2022){Prichard}, {Rafelski}, {Cooke}, {Me{\v{s}}tri{\'c}}, {Bassett}, {Ryan-Weber}, {Sunnquist}, {Alavi}, {Hathi}, {Wang}, {Revalski}, {Bajaj}, {O'Meara}, \& {Spitler}}]{prichard2022}
{Prichard}, L.~J., {Rafelski}, M., {Cooke}, J., {et~al.} 2022, \apj, 924, 14, \dodoi{10.3847/1538-4357/ac3004}

\bibitem[{{Rauch} {et~al.}(2011){Rauch}, {Becker}, {Haehnelt}, {Gauthier}, {Ravindranath}, \& {Sargent}}]{rauch2011}
{Rauch}, M., {Becker}, G.~D., {Haehnelt}, M.~G., {et~al.} 2011, \mnras, 418, 1115, \dodoi{10.1111/j.1365-2966.2011.19556.x}

\bibitem[{{Rivera-Thorsen} {et~al.}(2022){Rivera-Thorsen}, {Hayes}, \& {Melinder}}]{thorsen2022}
{Rivera-Thorsen}, T.~E., {Hayes}, M., \& {Melinder}, J. 2022, arXiv e-prints, arXiv:2206.10799.
\newblock \doarXiv{2206.10799}

\bibitem[{{Robertson} {et~al.}(2015){Robertson}, {Ellis}, {Furlanetto}, \& {Dunlop}}]{robertson2015}
{Robertson}, B.~E., {Ellis}, R.~S., {Furlanetto}, S.~R., \& {Dunlop}, J.~S. 2015, \apjl, 802, L19, \dodoi{10.1088/2041-8205/802/2/L19}

\bibitem[{{Saldana-Lopez} {et~al.}(2023){Saldana-Lopez}, {Schaerer}, {Chisholm}, {Calabr{\`o}}, {Pentericci}, {Cullen}, {Saxena}, {Amor{\'\i}n}, {Carnall}, {Fontanot}, {Fynbo}, {Guaita}, {Hathi}, {Hibon}, {Ji}, {McLeod}, {Pompei}, \& {Zamorani}}]{lopez2023}
{Saldana-Lopez}, A., {Schaerer}, D., {Chisholm}, J., {et~al.} 2023, \mnras, 522, 6295, \dodoi{10.1093/mnras/stad1283}

\bibitem[{{Salpeter}(1955)}]{salpeter1955}
{Salpeter}, E.~E. 1955, \apj, 121, 161, \dodoi{10.1086/145971}

\bibitem[{{Saxena} {et~al.}(2022){Saxena}, {Pentericci}, {Ellis}, {Guaita}, {Calabr{\`o}}, {Schaerer}, {Vanzella}, {Amor{\'\i}n}, {Bolzonella}, {Castellano}, {Fontanot}, {Hathi}, {Hibon}, {Llerena}, {Mannucci}, {Saldana-Lopez}, {Talia}, \& {Zamorani}}]{saxena2022}
{Saxena}, A., {Pentericci}, L., {Ellis}, R.~S., {et~al.} 2022, \mnras, 511, 120, \dodoi{10.1093/mnras/stab3728}

\bibitem[{{Shapley} {et~al.}(2016){Shapley}, {Steidel}, {Strom}, {Bogosavljevi{\'c}}, {Reddy}, {Siana}, {Mostardi}, \& {Rudie}}]{shapley2016}
{Shapley}, A.~E., {Steidel}, C.~C., {Strom}, A.~L., {et~al.} 2016, \apjl, 826, L24, \dodoi{10.3847/2041-8205/826/2/L24}

\bibitem[{{Shibuya} {et~al.}(2015){Shibuya}, {Ouchi}, \& {Harikane}}]{shibuya2015}
{Shibuya}, T., {Ouchi}, M., \& {Harikane}, Y. 2015, \apjs, 219, 15, \dodoi{10.1088/0067-0049/219/2/15}

\bibitem[{{Siana} {et~al.}(2015){Siana}, {Shapley}, {Kulas}, {Nestor}, {Steidel}, {Teplitz}, {Alavi}, {Brown}, {Conselice}, {Ferguson}, {Dickinson}, {Giavalisco}, {Colbert}, {Bridge}, {Gardner}, \& {de Mello}}]{siana2015}
{Siana}, B., {Shapley}, A.~E., {Kulas}, K.~R., {et~al.} 2015, \apj, 804, 17, \dodoi{10.1088/0004-637X/804/1/17}

\bibitem[{{Skelton} {et~al.}(2014){Skelton}, {Whitaker}, {Momcheva}, {Brammer}, {van Dokkum}, {Labb{\'e}}, {Franx}, {van der Wel}, {Bezanson}, {Da Cunha}, {Fumagalli}, {F{\"o}rster Schreiber}, {Kriek}, {Leja}, {Lundgren}, {Magee}, {Marchesini}, {Maseda}, {Nelson}, {Oesch}, {Pacifici}, {Patel}, {Price}, {Rix}, {Tal}, {Wake}, \& {Wuyts}}]{skelton2014}
{Skelton}, R.~E., {Whitaker}, K.~E., {Momcheva}, I.~G., {et~al.} 2014, \apjs, 214, 24, \dodoi{10.1088/0067-0049/214/2/24}

\bibitem[{{Sommariva} {et~al.}(2012){Sommariva}, {Mannucci}, {Cresci}, {Maiolino}, {Marconi}, {Nagao}, {Baroni}, \& {Grazian}}]{sommariva2012}
{Sommariva}, V., {Mannucci}, F., {Cresci}, G., {et~al.} 2012, \aap, 539, A136, \dodoi{10.1051/0004-6361/201118134}

\bibitem[{{Stark} {et~al.}(2013){Stark}, {Schenker}, {Ellis}, {Robertson}, {McLure}, \& {Dunlop}}]{stark2013}
{Stark}, D.~P., {Schenker}, M.~A., {Ellis}, R., {et~al.} 2013, \apj, 763, 129, \dodoi{10.1088/0004-637X/763/2/129}

\bibitem[{{Steidel} {et~al.}(2018){Steidel}, {Bogosavljevi{\'c}}, {Shapley}, {Reddy}, {Rudie}, {Pettini}, {Trainor}, \& {Strom}}]{steidel2018}
{Steidel}, C.~C., {Bogosavljevi{\'c}}, M., {Shapley}, A.~E., {et~al.} 2018, \apj, 869, 123, \dodoi{10.3847/1538-4357/aaed28}

\bibitem[{{Steidel} {et~al.}(2001){Steidel}, {Pettini}, \& {Adelberger}}]{steidel2001}
{Steidel}, C.~C., {Pettini}, M., \& {Adelberger}, K.~L. 2001, \apj, 546, 665, \dodoi{10.1086/318323}

\bibitem[{{Troncoso} {et~al.}(2014){Troncoso}, {Maiolino}, {Sommariva}, {Cresci}, {Mannucci}, {Marconi}, {Meneghetti}, {Grazian}, {Cimatti}, {Fontana}, {Nagao}, \& {Pentericci}}]{troncoso2014}
{Troncoso}, P., {Maiolino}, R., {Sommariva}, V., {et~al.} 2014, \aap, 563, A58, \dodoi{10.1051/0004-6361/201322099}

\bibitem[{{Urrutia} {et~al.}(2019){Urrutia}, {Wisotzki}, {Kerutt}, {Schmidt}, {Herenz}, {Klar}, {Saust}, {Werhahn}, {Diener}, {Caruana}, {Krajnovi{\'c}}, {Bacon}, {Boogaard}, {Brinchmann}, {Enke}, {Maseda}, {Nanayakkara}, {Richard}, {Steinmetz}, \& {Weilbacher}}]{urrutia2019}
{Urrutia}, T., {Wisotzki}, L., {Kerutt}, J., {et~al.} 2019, \aap, 624, A141, \dodoi{10.1051/0004-6361/201834656}

\bibitem[{{Vanzella} {et~al.}(2010){Vanzella}, {Siana}, {Cristiani}, \& {Nonino}}]{vanzella2010}
{Vanzella}, E., {Siana}, B., {Cristiani}, S., \& {Nonino}, M. 2010, \mnras, 404, 1672, \dodoi{10.1111/j.1365-2966.2010.16408.x}

\bibitem[{{Vanzella} {et~al.}(2012){Vanzella}, {Guo}, {Giavalisco}, {Grazian}, {Castellano}, {Cristiani}, {Dickinson}, {Fontana}, {Nonino}, {Giallongo}, {Pentericci}, {Galametz}, {Faber}, {Ferguson}, {Grogin}, {Koekemoer}, {Newman}, \& {Siana}}]{vanzella2012}
{Vanzella}, E., {Guo}, Y., {Giavalisco}, M., {et~al.} 2012, \apj, 751, 70, \dodoi{10.1088/0004-637X/751/1/70}

\bibitem[{{Vanzella} {et~al.}(2015){Vanzella}, {de Barros}, {Castellano}, {Grazian}, {Inoue}, {Schaerer}, {Guaita}, {Zamorani}, {Giavalisco}, {Siana}, {Pentericci}, {Giallongo}, {Fontana}, \& {Vignali}}]{vanzella2015}
{Vanzella}, E., {de Barros}, S., {Castellano}, M., {et~al.} 2015, \aap, 576, A116, \dodoi{10.1051/0004-6361/201525651}

\bibitem[{{Vanzella} {et~al.}(2016){Vanzella}, {de Barros}, {Vasei}, {Alavi}, {Giavalisco}, {Siana}, {Grazian}, {Hasinger}, {Suh}, {Cappelluti}, {Vito}, {Amorin}, {Balestra}, {Brusa}, {Calura}, {Castellano}, {Comastri}, {Fontana}, {Gilli}, {Mignoli}, {Pentericci}, {Vignali}, \& {Zamorani}}]{vanzella2016}
{Vanzella}, E., {de Barros}, S., {Vasei}, K., {et~al.} 2016, \apj, 825, 41, \dodoi{10.3847/0004-637X/825/1/41}

\bibitem[{{Vanzella} {et~al.}(2018){Vanzella}, {Nonino}, {Cupani}, {Castellano}, {Sani}, {Mignoli}, {Calura}, {Meneghetti}, {Gilli}, {Comastri}, {Mercurio}, {Caminha}, {Caputi}, {Rosati}, {Grillo}, {Cristiani}, {Balestra}, {Fontana}, \& {Giavalisco}}]{vanzella2018}
{Vanzella}, E., {Nonino}, M., {Cupani}, G., {et~al.} 2018, \mnras, 476, L15, \dodoi{10.1093/mnrasl/sly023}

\bibitem[{{Verhamme} {et~al.}(2015){Verhamme}, {Orlitov{\'a}}, {Schaerer}, \& {Hayes}}]{verhamme2015}
{Verhamme}, A., {Orlitov{\'a}}, I., {Schaerer}, D., \& {Hayes}, M. 2015, \aap, 578, A7, \dodoi{10.1051/0004-6361/201423978}

\bibitem[{{Wang} {et~al.}(2023){Wang}, {Teplitz}, {Smith}, {Windhorst}, {Rafelski}, {Mehta}, {Alavi}, {Brammer}, {Colbert}, {Grogin}, {Hathi}, {Koekemoer}, {Prichard}, {Scarlata}, {Sunnquist}, {Arrabal Haro}, {Conselice}, {Gawiser}, {Guo}, {Hayes}, {Jansen}, {Ji}, {Lucas}, {O'Connell}, {Robertson}, {Rutkowski}, {Siana}, {Vanzella}, {Ashcraft}, {Bagley}, {Baronchelli}, {Barro}, {Blanche}, {Broussard}, {Carleton}, {Chartab}, {Cheng}, {Codoreanu}, {Cohen}, {Dai}, {Darvish}, {Dave}, {DeGroot}, {De Mello}, {Dickinson}, {Emami}, {Ferguson}, {Ferreira}, {Finkelstein}, {Finkelstein}, {Gardner}, {Gburek}, {Giavalisco}, {Grazian}, {Gronwall}, {Hemmati}, {Howell}, {Iyer}, {Kaviraj}, {Kurczynski}, {Lazar}, {MacKenty}, {Mantha}, {Martin}, {Martin}, {McCabe}, {Mobasher}, {Nedkova}, {Olsen}, {Otteson}, {Ravindranath}, {Redshaw}, {Sattari}, {Soto}, {Yung}, {Zabelle}, \& {the UVCANDELS team}}]{wang2023}
{Wang}, X., {Teplitz}, H.~I., {Smith}, B.~M., {et~al.} 2023, arXiv e-prints, arXiv:2308.09064, \dodoi{10.48550/arXiv.2308.09064}

\bibitem[{{Whitaker} {et~al.}(2019){Whitaker}, {Ashas}, {Illingworth}, {Magee}, {Leja}, {Oesch}, {van Dokkum}, {Mowla}, {Bouwens}, {Franx}, {Holden}, {Labb{\'e}}, {Rafelski}, {Teplitz}, \& {Gonzalez}}]{whitaker2019}
{Whitaker}, K.~E., {Ashas}, M., {Illingworth}, G., {et~al.} 2019, \apjs, 244, 16, \dodoi{10.3847/1538-4365/ab3853}

\bibitem[{{Williams} {et~al.}(2023){Williams}, {Tacchella}, {Maseda}, {Robertson}, {Johnson}, {Willott}, {Eisenstein}, {Willmer}, {Ji}, {Hainline}, {Helton}, {Alberts}, {Baum}, {Bhatawdekar}, {Boyett}, {Bunker}, {Carniani}, {Charlot}, {Chevallard}, {Curtis-Lake}, {de Graaf}, {Egami}, {Franx}, {Kumari}, {Maiolino}, {Nelson}, {Rieke}, {Sandles}, {Shivaei}, {Simmonds}, {Smit}, {Suess}, {Sun}, {Ubler}, \& {Witstok}}]{williams2023}
{Williams}, C.~C., {Tacchella}, S., {Maseda}, M.~V., {et~al.} 2023, arXiv e-prints, arXiv:2301.09780, \dodoi{10.48550/arXiv.2301.09780}

\bibitem[{{Windhorst} {et~al.}(2011){Windhorst}, {Cohen}, {Hathi}, {McCarthy}, {Ryan}, {Yan}, {Baldry}, {Driver}, {Frogel}, {Hill}, {Kelvin}, {Koekemoer}, {Mechtley}, {O'Connell}, {Robotham}, {Rutkowski}, {Seibert}, {Straughn}, {Tuffs}, {Balick}, {Bond}, {Bushouse}, {Calzetti}, {Crockett}, {Disney}, {Dopita}, {Hall}, {Holtzman}, {Kaviraj}, {Kimble}, {MacKenty}, {Mutchler}, {Paresce}, {Saha}, {Silk}, {Trauger}, {Walker}, {Whitmore}, \& {Young}}]{windhorst2011}
{Windhorst}, R.~A., {Cohen}, S.~H., {Hathi}, N.~P., {et~al.} 2011, \apjs, 193, 27, \dodoi{10.1088/0067-0049/193/2/27}

\bibitem[{{Wise} \& {Cen}(2009)}]{wise2009}
{Wise}, J.~H., \& {Cen}, R. 2009, \apj, 693, 984, \dodoi{10.1088/0004-637X/693/1/984}

\bibitem[{{Witten} {et~al.}(2024){Witten}, {Laporte}, {Martin-Alvarez}, {Sijacki}, {Yuan}, {Haehnelt}, {Baker}, {Dunlop}, {Ellis}, {Grogin}, {Illingworth}, {Katz}, {Koekemoer}, {Magee}, {Maiolino}, {McClymont}, {P{\'e}rez-Gonz{\'a}lez}, {Pusk{\'a}s}, {Roberts-Borsani}, {Santini}, \& {Simmonds}}]{witten2024}
{Witten}, C., {Laporte}, N., {Martin-Alvarez}, S., {et~al.} 2024, Nature Astronomy, \dodoi{10.1038/s41550-023-02179-3}

\bibitem[{{Yamada} {et~al.}(2012){Yamada}, {Matsuda}, {Kousai}, {Hayashino}, {Morimoto}, \& {Umemura}}]{yamada2012}
{Yamada}, T., {Matsuda}, Y., {Kousai}, K., {et~al.} 2012, \apj, 751, 29, \dodoi{10.1088/0004-637X/751/1/29}

\bibitem[{{Yuan} {et~al.}(2021){Yuan}, {Zheng}, {Lin}, {Zhu}, \& {Rahna}}]{yuan2021}
{Yuan}, F.-T., {Zheng}, Z.-Y., {Lin}, R., {Zhu}, S., \& {Rahna}, P.~T. 2021, \apjl, 923, L28, \dodoi{10.3847/2041-8213/ac4170}

\end{thebibliography}
\bibliographystyle{aasjournal}

\end{document}